\documentclass[12pt]{iopart}

\usepackage[numbers,sort&compress]{natbib}
\usepackage{iopams}

\expandafter\let\csname equation*\endcsname\relax

\expandafter\let\csname endequation*\endcsname\relax
\usepackage[cmex10]{amsmath}

\usepackage{algorithmic}
\usepackage{array}
\usepackage[caption=false,font=footnotesize]{subfig}
\usepackage{fixltx2e}
\usepackage{stfloats}
\usepackage{graphicx}
\usepackage{multirow}
\usepackage{threeparttable}

\begin{document}
\title[ERASE for removal of EMG artifacts]{Electromyogram (EMG) Removal by Adding Sources of EMG (ERASE) - A novel ICA-based algorithm for removing myoelectric artifacts from EEG - Part 1}

\author{Yongcheng Li$^1$, Po T. Wang$^2$,
Mukta P. Vaidya$^{3,4,5}$, Charles Y. Liu $^{6,7,8}$, Marc W. Slutzky $^{3,4,5}$ and An H. Do$^1$}

\address{$^1$ Department of Neurology, University of California, Irvine, CA 92697 USA}
\address{$^2$ Department of Biomedical Engineering, University of California, Irvine, CA 92697, USA}
\address{$^3$ Department of Neurology, Northwestern University, Chicago, Illinois, USA}
\address{$^4$ Department of Physiology, Northwestern University, Chicago, Illinois, USA}
\address{$^5$ Department of Physical Medicine and Rehabilitation, Northwestern University, Chicago, Illinois, USA}
\address{$^6$ Department of Neurosurgery, University of Southern California, CA, USA}
\address{$^7$ Rancho Los Amigos National Rehabilitation Center, CA, USA}
\address{$^8$ Neurorestoration Center, University of Southern California, CA, USA}

\ead{and@uci.edu; yongchel@uci.edu}

\begin{abstract}
Electroencephalographic (EEG) recordings are often contaminated by electromyographic (EMG) artifacts, especially when recording during movement. Existing methods to remove EMG artifacts include independent component analysis (ICA), and other high-order statistical methods. However, these methods can not effectively remove most of EMG artifacts. Here, we proposed a modified ICA model for EMG artifacts removal in the EEG, which is called EMG Removal by Adding Sources of EMG (ERASE). In this new approach, additional channels of real EMG from neck and head muscles (reference artifacts) were added as inputs to ICA in order to ``force'' the most power from EMG artifacts into a few independent components (ICs). The ICs containing EMG artifacts (the ``artifact ICs'') were identified and rejected using an automated procedure. ERASE was validated first using both simulated and experimentally-recorded EEG and EMG.  Simulation results showed ERASE removed EMG artifacts from EEG significantly more effectively than conventional ICA. Also, it had a low false positive rate and high sensitivity. Subsequently, EEG was collected from 8 healthy participants while they moved their hands to test the realistic efficacy of this approach. Results showed that ERASE successfully removed EMG artifacts (on average, about 75\% of EMG artifacts were removed when using real EMGs as reference artifacts) while preserving the expected EEG features related to movement. We also tested the ERASE procedure using simulated EMGs as reference artifacts (about 63\% of EMG artifacts removed). Compared to conventional ICA, ERASE removed on average 26\% more EMG artifacts from EEG. These findings suggest that ERASE can achieve significant separation of EEG signal and EMG artifacts without a loss of the underlying EEG features. These results indicate that using additional real or simulated EMG sources can increase the effectiveness of ICA in removing EMG artifacts from EEG. Combined with automated artifact IC rejection, ERASE also minimizes potential user bias. Future work will focus on improving ERASE so that it can also be used in real-time applications.
\end{abstract}
\noindent{artifacts removal, ICA, electroencephalogram (EEG), muscle artifacts, neural network}
\maketitle
\normalsize

\section{Introduction}

Electroencephalographic (EEG) signals are often contaminated by surface electromyographic (EMG) signals and can make it difficult to appropriately interpret EEG signals or use them in various neuroengineering applications. There are many approaches to remove EMG artifacts from EEG signals, which can be broadly classified in the following categories \cite{11,43,a1,a2}: 1). filtering (e.g. adaptive filtering, Wiener filtering, and Bayes filtering \cite{a3,a4,a5}), 2). linear regression method \cite{a6,a7,a8}, 3). blind source separation (BSS) methods which includes principal component analysis (PCA) \cite{a9,a10}, independent component analysis (ICA) \cite{a11,a12,a13,a14}, canonical correlation analysis (CCA) \cite{a15,a16,56,a18}, sparse component analysis (SCA) \cite{a19}, singular spectrum analysis (SSA) \cite{a20}, and etc. 4). source decomposition, including discrete wavelet transform (DWT) \cite{a15,a21,a22,a23}, empirical mode decomposition (EMD) \cite{a24,a25,58}, stationary wavelet transform (SWT) \cite{a27,a28} and ensemble empirical mode decomposition (EEMD) \cite{a29,a30,a18,a32}, 5). neural networks (NN) \cite{a33,a35} and adaptive neural fuzzy inference systems (ANFIS) \cite{a34,a36}. In the setting of a plethora of EMG artifacts removal algorithms, previous surveys \cite{11,a1} suggest that BSS methods are the most commonly used and outperform other approaches at EMG artifacts removal from EEG.
Among the BSS methods mentioned above, Urigüen et al \cite{a1} states that 45\% of work in their bibliography use ICA to remove the EMG artifact from EEG. CCA is another widely employed method for the removal of EMG artifacts from EEG in recent years \cite{56,57}. However, it does not outperform ICA at removing EMG artifacts from EEG \cite{a15,a37,a38} or at removing ocular and cardiac artifacts \cite{a39,a40,a41,a42,a43,a44}.

Despite the popularity of ICA for EMG artifact rejection in EEG, its use is still affected by several issues \cite{8,14,11}. For example, nearly all EEG channels are typically contaminated by EMG, and there is a high spatiotemporal overlap between EMG artifacts and EEG signal \cite{30}. Therefore, conventional ICA algorithms are usually unable to separate EMG artifacts from the EEG signal---that is, it is difficult to ``force'' all of the EMG artifacts into an isolated set of independent components. Hence, post-ICA-treated data may still include residual EMG \cite{30,46}. Also, since the EMGs could have multiple source locations and show a large overlap with the higher frequency components ($>$20 Hz) of EEG signals, it is difficult to assign a universal operational definition for EMG components \cite{47,48}. Hence, rejection of EMG artifact components is typically performed manually \emph{when no prior knowledge about artifacts is available}. This leads to potential over- or under-rejection of components as users attempt to distinguish between neurogenic and myogenic components in common ICA. Further, rejecting components manually is cumbersome/time-consuming and can introduce subjectivity. Additionally, accurate extraction of source signals in the ICA model is another issue since the global optimum in these algorithms is typically affected by the contrast function. Some approaches, which are presented in further detail below, were developed to solve these issues.

 Some studies demonstrated that prior knowledge about artifacts or source signals can improve the effectiveness of ICA at removing artifacts \cite{a1,a49}. Therefore, constrained ICA (cICA) or ICA with reference (ICA-R) which incorporate prior knowledge about the artifact and/or source signals were developed \cite{a11}. This is performed by imposing temporal or spatial constraints on the source mixture model. In temporally constrained ICA, prior knowledge about artifacts can be introduced into the ICA model to identify the artifact IC/ICs by solving a constrained optimization problem \cite{a11,a39,a45,13,a47}.
 Temporally constrained ICA can only be used for the removal of EOG and ECG, but is not highly effective for EMG artifacts \cite{13,a47,a39}. EMG artifacts sources, which are more time-varying and non-stationarity, are too complicated to characterize for optimization constraints. Hence, no prior studies have adequately addressed the separation of spatiotemporal overlap between EMG and EEG by using temporally constrained ICA. Spatially constrained ICA incorporates prior knowledge or assumption of spatial topographies of some source projections acting as a spatial filter, and limit the degree to which some of the columns of the mixing matrix may deviate from the known projections \cite{a48,a49,a10,a51}.
 As mentioned above, EMG artifacts are time-varying and highly overlapped with EEG. Hence, known spatial topographies of EMG source projections derived from previous data recording or mathematically simulated model are usually inaccurate and difficult to achieve. Although some work on EMG artifact removal utilized spatially constrained ICA with EEG during seizures \cite{a48}, the performance of spatially constrained ICA on these ictal EEG, where the ground truth of EMG artifacts is actually unknown, has not been fully and rigorously established. Hence, using spatially constrained ICA to remove the EMG artifacts from real EEG is still unsubstantiated. Additionally, running ICA iteratively is required for both types of constrained ICA to reject the artifacts, which is time-consuming and hard to achieve real-time application. Even so, when the prior knowledge about artifacts or source signal is available, some form of spatially constrained ICA is preferred as compared to using ICA alone \cite{a1,a49}.

Another important issue associated with removing EMG artifacts from EEG via ICA is automated EMG artifact rejection. Generally, \emph{when reference waveforms are available,} there is one method to achieve automated rejection based on ICA in previous studies \cite{a1}. Specifically, cICA compares spatial and temporal statistical characteristics of ICs to those from background EEG or artifacts. Subsequently, a combination of thresholds for those statistical characteristics is used in an automated algorithm to reject the artifacts \cite{a14,a12,a52,7,8}. This method has been demonstrated to reliably remove EOG and ECG artifacts, particularly since these signals drastically differ from EEG in spatial and temporal statistical characteristics \cite{a39,a40,a53}. However, using this method for EMG artifact removal is still inadequate due to the spatiotemporal overlap between EMG artifacts and EEG signals (i.e. EMG artifacts always introduce a large number of unique scalp maps, leaving few ICs available for capturing brain sources). Therefore, it is necessary to develop an automated technique that can more effectively and systematically remove EMG artifacts while not affecting any of the underlying signal features. 

In order to improve the effectiveness and reliability of ICA in removing the EMG artifacts from EEG and establish an effective automated artifacts rejection procedure, we introduce a novel method, termed as \textit{EMG Removal by Adding Sources of EMG} (ERASE) and subject it to rigorous mathematical and experimental validation. ERASE combines the advantages of two types of cICA. It aims to improve upon ICA by adding either real or simulated EMG artifacts as extra ``reference'' channel signals into the EEG data. We mathematically demonstrated that if the reference EMG artifacts were not independent of the contaminant EMG artifacts in EEG, a larger proportion of the artifacts could be identified by several specific independent components (ICs) after running ICA. Also, we proposed criteria based on the mixing matrix to automatically identify and reject the artifacts components. Our results revealed that ERASE had higher effectiveness in removing the EMG artifacts compared to conventional ICA. In summary, this study developed an effective EMG rejection approach, which can provide more confidence for the utilization of EEG in applications such as physiological studies underlying motor behaviors.

\section{Methods}


\subsection{Description of ERASE ICA model based on added EMG sources}
\subsubsection{Model description}
To facilitate more effective removal of EMG artifacts from the EEG data, we combined EMG artifacts (here, either simulated EMG or recorded EMG) with EEG datasets and applied a modified ICA model as follows:
\begin{eqnarray}\label{1}
   {{\hat{X}_{t}}\choose{n^{\ast}_{\tau}}} &= A_{t+ \tau} \times S_{t+ \tau} \nonumber\\
   {{X_{t}+b_{t}\cdot N_{t}}\choose{n^{\ast}_{\tau}}} &= A_{t+\tau}\times {{s_{t}+m_{t}}\choose{m^{\ast}_{\tau}}}
\end{eqnarray}
where $t$ is the number of the EEG channels ($t$ dimension), and $\tau$ is the number of the reference EMG channels ($\tau$ dimension), $\hat{X_{t}}=X_{t}+b_{t}\cdot N_{t}$, and $X_{t}$ is the uncontaminated EEG data; $N_{t}$ is the contaminant EMG artifacts, which usually is the real EMG artifacts inside of EEG; $b_{t}$ is the linear coefficients; $n^{\ast}_{\tau}$ is the reference EMG artifacts, which are extra channels containing EMG signals from muscles or simulation; $A_{t+\tau}$ is the mixing matrix of dimension $(t+\tau) \times (t+\tau)$, $S_{t+\tau}$ is the independent component sources with $t+\tau$ dimension, in which $s_{t}$ is the sources representing the uncontaminated EEG, $m_{t}$ are the sources representing the contaminant EMG artifacts, $m^{\ast}_{\tau}$ are the reference EMG sources.

\textbf{Theorem:} Given that the reference EMG sources are independent, and $n^{\ast}_{\tau}$ is only dependent with $N_{t}$, then $m_{t}=0$.

\textbf{Proof:} We assume that $m_{t}\neq0$. In this model, $N_{t}=a1_{t} \times m_{t}+ a2_{\tau}\times m^{\ast}_{\tau}$ and $n^{\ast}_{\tau}=a3_{t} \times m_{t}+ a4_{\tau}\times m^{\ast}_{\tau}$ ($a1$, $a2$, $a3$ and $a4$ are the corresponding submatrices of the mixing matrix). Since $N_{t}$ is dependent with $n^{\ast}_{\tau}$, a relationship can be defined as follows:
\begin{eqnarray}\label{2}
 \mathbf{Y:}=(y(N_{t}(1)),y(N_{t}(2)),...,y(N_{t}(m)),y(n^{\ast}_{\tau}(1)),&y(n^{\ast}_{\tau}(2)),...,y(n^{\ast}_{\tau}(n)))\nonumber\\ &(m \leq t, n \leq \tau)
\end{eqnarray}
where $y(.)$ is probability density function.\\
This expression of $N_{t}$ and $n^{\ast}_{\tau}$ can be combined as follows:

\begin{eqnarray}\label{3}
  \mathbf{Y:}=&(y(a1_{t} \times m_{t}(1)+ a2_{\tau}\times m^{\ast}_{\tau}(1)), y(a1_{t} \times m_{t}(2)+ a2_{\tau}\times m^{\ast}_{\tau}(2))\nonumber\\&,...,y(a1_{t} \times m_{t}(m)+ a2_{\tau}\times m^{\ast}_{\tau}(m)), \nonumber\\
  &y(a3_{t} \times m_{t}(1)+ a4_{\tau}\times m^{\ast}_{\tau}(1)), y(a3_{t} \times m_{t}(2)+ a4_{\tau}\times m^{\ast}_{\tau}(2))\nonumber\\&,...,y(a3_{t} \times m_{t}(n)+ a4_{\tau}\times m^{\ast}_{\tau}(n)))\qquad (m \leq t, n \leq \tau)
\end{eqnarray}

Given that: $z(m_{t},m^{\ast}_{\tau})=y(a_{t} \times m_{t}+ a_{\tau}\times m^{\ast}_{\tau})$. We get:
\begin{eqnarray}\label{4}
\mathbf{Z:}= (z(m_{t}(1)),z(m_{t}(2)),...,z(m_{t}(m)),z(m^{\ast}_{\tau}(1)),&z(m^{\ast}_{\tau}(2)),...,
z(m^{\ast}_{\tau}(n)))\nonumber\\&(m \leq t, n \leq \tau)
\end{eqnarray}
where $z(.)$ is another expression of $y(.)$, $N_{t}(m)$ is the contaminant artifacts in the $m^{th}$ channel, $n^{\ast}_{\tau}(n)$ is the reference EMG artifacts in the $n^{th}$ channel, $m_{t}(i)$ is the $i^{th}$ sources representing the contaminant EMG artifacts, and $m^{\ast}_{\tau}(j)$ is the $j^{th}$ reference EMG sources. From the equation (4), we know that $m_{t}$ is dependent with $m^{\ast}_{\tau}$. This denotes that the reference EMG source are dependent with sources representing contaminant EMG artifacts. Since this violates the ICA principle of component independence, $m_{t}$ must equal 0.

Two key assumptions are made in our model. One is the independence among the reference EMG sources and the other was the dependence between the contaminant EMG artifacts and the reference EMG artifacts. This means if the simulated EMG can meet these two assumptions, in situations where real EMG was not collected, or the situation does not allow for EMG recordings, simulated EMG can also act as the reference EMG artifacts.

\subsubsection{Rejection criteria}
We define the $(t+\tau) \times (t+\tau)$ mixing matrix $A_{t+\tau}$ in equation (1) as below:
\begin{eqnarray}\label{5}
A_{t+\tau}={
\left(
\begin{matrix}
 a_{1,1}      & a_{1,2}      & \cdots & a_{1,t+\tau}      \\
 a_{2,1}      & a_{2,2}      & \cdots & a_{2,t+\tau}      \\
 \vdots & \vdots & \ddots & \vdots \\
 a_{t,1}      & a_{t,2}      & \cdots & a_{t,t+\tau}      \\
 a_{t+1,1}      & a_{t+1,2}      & \cdots & a_{t+1,t+\tau}\\
 \vdots & \vdots & \ddots & \vdots \\
 a_{t+\tau,1}      & a_{t+\tau,2}      & \cdots & a_{t+\tau,t+\tau}\\
\end{matrix}
\right)
}
\end{eqnarray}
where the first $t$ rows are the coefficients corresponding to the EEG channels, the last $\tau$ rows are the coefficients corresponding to the reference EMG channels, the coefficients in each column represent the weights of this IC to all EEG/EMG channels.

To develop an automatic method of identifying and rejecting the independent components related to EMG artifacts (referred to as ``artifact ICs'') in real EEG data after running ICA, two criteria were defined:
\begin{itemize}
  \item First, a threshold was established based on the root mean square (RMS) values of coefficients in the mixing matrix rows corresponding to the EMG channels, which was defined as:
\begin{eqnarray}\label{6}
      R_{ms}=\frac{\sum(\sqrt{\frac{\sum_{n=1}^{t+\tau}a_{t+1,n}^{2}}{t+\tau}}+...+\sqrt{\frac{\sum_{n=1}^{t+\tau}a_{t+\tau,n}^{2}}{t+\tau}})}{\tau}
\end{eqnarray}
      where $R_{ms}$ denotes the RMS value of coefficients. The true threshold was calculated by RMS value times gain. Gain is a constant which was empirically set between 0.4--3. \emph{The ICs whose absolute value of coefficients in the corresponding EMG rows were above the threshold were defined to be artifact ICs.}
  \item Second, the ICs whose maximal absolute value of coefficients ($max(|a_{1,i}|,...,|a_{t+\tau,i}|),1\leq i \leq (t+\tau)$) corresponds to a hat band electrode, were rejected. Note that hat band electrodes were defined as all the EEG electrodes which were on the outermost circumference of the head, as defined by \cite{54} (Supplementary Fig. 8).
\end{itemize}
In order to find the proper threshold, we changed the gain set with 0.1 intervals so that a threshold set was limited to 0.4--3 times RMS value. The threshold was automatically set at the value which simultaneously minimizes high-frequency band (40 -- 100 Hz) synchronization for all the EEG channels and maximizes $\mu$ (8 -- 12 Hz) desynchronization in the EEG channel of interest (e.g. C3/C4 for hand movements). Specifically, the threshold was decided automatically by finding the minimal value from the summated high-frequency band synchronization and $\mu$ band desynchronization vector (an example of finding the proper gain was shown in Fig. 2.).

\subsection{Validation with Simulated EEG/EMG Data}
 To mathematically verify ERASE, simulated EMG and EEG were generated, subjected to ERASE, and its performance was assessed by several metrics as follows. Here, simulated EMG was also used as reference EMG artifacts for the experimental data processing.
\subsubsection{Generating simulated EMG} \label{secsimulatedEMG}
The simulated EMG was generated as the contaminant and reference EMG artifacts, using the approach below \cite{40}:
\begin{enumerate}
  \item The Hodgkin-Huxley model was used to simulate extracellular current. For skeletal muscles, the Hodgkin-Huxley model is a widely accepted model for simulating extracellular current \cite{41}.
  \item Single fiber action potentials (SFAP) were generated with a volume conduction model, defined as follows \cite{9}:
  \begin{eqnarray}\label{7}
  \fl V_{E}(z,y) = K\{\int_{S_{1}}\frac{\partial e(z)}{\partial z}\cdot \frac{1}{r}dS+\int_{S}dS\int_{-\infty}^{+\infty}\frac{\partial^2e(z)}{\partial z^2} \cdot \frac{1}{r}dz -\int_{S_{2}}\frac{\partial e(z)}{\partial z}\cdot \frac{1}{r}dS\}
\end{eqnarray}
where $V_{E}$ is the SFAP, $e(z)$ is the extracellular current (from step 1 above), $z$ and $y$ are the axial and radial directions, respectively, $S_{1}$ and $S_{2}$ are the fiber sections at the fiber ends, and $r$ is the distance between the surface element, and $dS$ is the observation point. The equation above was discretized to generate the SFAP using known parameter values from the literature, including fiber length, endplate position, observation position, etc. \cite{10,11}.
  \item A Gaussian distribution with 0 mean and standard deviation (SD) = 2.5 mm \cite{10} was used to depict the endplate positions. The voltage propagation velocities were considered as a Gaussian distribution with an average of 4 m/s and SD = 0.125 \cite{10}. A total of 100 SFAPs were first generated and their average served as one activation of the motor unit action potential (MUAP).
  \item A Poisson process was employed to model the firing rate of the MUAPs (as defined in \cite{10}). The EMG firing rate and amplitude were assumed to increase with the hand/finger movements. Hence, different firing rates were applied to different phases (idle vs movement). For each muscle, the new Poisson process with the same firing rate was launched to generate the time points of the firing of MUAP.
  \item Eight different facial muscles, including bilateral frontalis, temporalis, masseter, and trapezius were simulated for each session (one session denoted one record, which included several trials). Each muscle's simulated EMG was filtered based on its frequency characteristics as described in the literature \cite{11}.
\end{enumerate}

The above approach ensured that the simulated reference EMG was dependent on the contaminant EMG artifacts to some degree. Therefore, the effectiveness in removing EMG artifacts by using the simulated EMG should be similar with that obtained by using the real EMG.

These simulated EMG artifacts were then combined with the simulated EEG data (as described below) as separate channels (the ``EMG channels''). These separate channels were located at different positions on the edge of brain topographic map (Supplementary Fig. 8). The coordinates of these positions corresponded to the approximate muscle locations on the head.

\subsubsection{Generating simulated EEG}

Simulated EEG data was generated by using the approach in \cite{31}:
\begin{enumerate}
  \item Simulated EEG was created using a linear mixture of five Gaussian noises. Each Gaussian noise was bandpass filtered in different frequency bands (1-30 Hz, 20-40 Hz, 40-80 Hz, 80-100 Hz, and 100-200 Hz). The amplitude and variance of each Gaussian noise were adjusted to fit the average values of real EEG data.
  \item To mimic the spatial correlation between EEG channels due to volume conduction, a smoothing convolution was performed across channels to increase the spatial correlation amongst adjacent channels. The smoothing convolution kernel was a Gaussian function with a standard deviation equal to 4 channels. We considered the last channels on the list were correlated with the first channels on the list. Note that this resulted in consecutive channels being highly correlated (i.e. channel 1 becomes highly correlated to channel 2, channel 2 to 3, etc).
\end{enumerate}

Simulated EEG data sets were generated with a temporal structure similar to real EEG. 
 More specifically, each simulated dataset contained 32 channels, the maximal amplitude was set to 60 $\mu$V with a variance of 30 $\mu$$V^2$, and the sampling rate was set to 2000 Hz. A total of 5 minutes of simulated EEG was generated. Theoretically, the number of EEG channels (still meet the minimum for ICA run) has no effect on our approach, so we can select any number of EEG channels as long as we feel it is appropriate.


\subsubsection{Performance Assessment}
The performance of ERASE was assessed in two simulated scenarios. In Scenario 1, ERASE was tested across an increasing number of contaminated EEG channels. Here, 3 types of simulated EMG (frontalis, temporalis, and masseter) were generated (as in Section 2.2.1). In the first iteration, 200 sets of simulated EEG (32 channels) were generated (6 channels were contaminated). Specifically, pairs of simulated EEG channels were then contaminated by a single simulated EMG type, and this was repeated until all 3 types were exhausted. Simulated EMG was multiplied by a randomly generated weight factor and added to the simulated EEG. For each simulated EMG, 2 normally distributed pseudorandom numbers were generated (mean: 0 and standard deviation: 1) and normalized to act as weight factors. Subsequently, the simulated EMG without multiplication was combined with the contaminated EEG set as separate channels. This combined dataset was subjected to ICA. This process was repeated with 12, 18, 24, and 30 simulated EEG channels, in which groups of 4, 6, 8, and 10 simulated EEG channels, respectively, were contaminated with a single EMG type.

 For Scenario 2, ERASE was tested on simulated EEG across an increasing burden of EMG contamination. Here, 1 set of simulated EMG was generated for each of the following types: frontalis, temporalis, masseter, trapezius, and eye blinks (as in Section 2.2.1).  Also, 200 sets of simulated EEG were generated as described above. For each set of simulated EEG, 6 randomly chosen simulated EEG channels were contaminated by a single EMG type. More specifically, the simulated EMG was multiplied by a randomly generated weight factor and then added to the simulated EEG. For each simulated EMG, 6 normally distributed pseudorandom numbers were generated (mean: 0 and standard deviation: 1) and normalized to act as weight factors. Likewise, the simulated EMG without multiplication were combined with the contaminated set of EEG as separate channels, thereby acting as reference EMG artifacts (e.g. there were 32 EEG channels, the simulated EMG was the 33rd, 34th,...37th channel, depending on how many simulated EMG channels were used). This combined dataset was subjected to ICA. This process was repeated, each time incrementally adding another simulated EMG type to an additional 6 randomly chosen simulated EEG channels, until all EMG types were exhausted. Note that to simplify the process, one simulated EEG channel was contaminated with no more than one type of simulated EMG in both scenarios.

Each combined simulated EEG/EMG set was then subjected to the ICA algorithm (FastICA). The performance was assessed by calculating the effectiveness, false positive rate and sensitivity (described below) across all of the above data sets. 
Note that when using simulated EEG (Section 2.2.2), a threshold was not necessary to define artifact ICs. Instead, for each reference EMG channel, the IC with the highest coefficient was defined as an artifact IC.

\textbf{Effectiveness:}
To compare how well EMG artifacts were removed between ERASE and the conventional ICA, the effectiveness of both methods were compared. Here, effectiveness was defined as the ratio of the amount of simulated EEG signal in the artifact ICs. Effectiveness was expressed as artifact index, which was defined as the following:
\begin{eqnarray}\label{8}
AI=\frac{\sum(|a_{i}^{*}|+...+|a_{i+j}^{*}|)/j}{\sum(|a_{1}|+...+|a_{i-1}|+|a_{i+j+1}|+...+|a_{32}|)/(32-j)}
\end{eqnarray}
where 
 $a$ and $a^{*}$ are defined in equation (9). Equation (8) denotes the ratio of the average mixing matrix coefficient values in the contaminated rows and those in the uncontaminated rows in the artifacts IC columns (referred to equation (9)).
A larger artifact index indicated that this identified artifact IC contained more artifacts, but less simulated EEG signal.

We applied ERASE to two conditions of EMG-artifacts contaminated simulated EEG: with and without separate channels. Generally, we referred to this latter case as ``conventional ICA'' condition. The artifact indices calculated for the conditions with and without simulated EMG were statistically compared in order to verify the effectiveness of our model. Note that only criteria 2 from the rejection criteria above was employed in the conventional ICA condition.

\textbf{False positive rate:}
To determine how frequently ERASE would erroneously detect sources that were independent of the reference EMG artifacts, a false positive rate was designed.
More specifically, false positive rate determined how often contaminant artifacts ($N_{t}$, equation 1) that were \emph{independent of} the reference EMG artifacts ($n^{\ast}_{\tau}$, equation 1) were erroneously ``pushed'' into the artifact ICs. To assess this, $N_{t}$, composed of Gaussian random noise (mean 0, S.D. 30), was used to further contaminate the simulated EEG. Simulated EMG was still used as reference EMG artifacts, $n^{\ast}_{\tau}$,  and combined with the contaminated EEG data sets and acted as separate channels as described above. To simplify the process, only one type of (independent) Gaussian random artifacts was employed for both scenarios.

Here, we define the mixing matrix column corresponding to the artifact IC (the ``artifact IC column'') as follows:
\begin{eqnarray}\label{9}
 \fl V=(a_{1},...,a_{i-1},a_{i}^{*},...,a_{i+j}^{*},a_{i+j+1},...,a_{32},\tilde{a}_{33},...,\tilde{a}_{36})^{T} (1 \leq i \leq 26, 6 \leq i+j \leq 32)
\end{eqnarray}
 where $V$ is the ``artifact IC column'', $a$ denotes the coefficients corresponding to uncontaminated channels (the ``non-artifacts coefficients''), $a^{*}$ denotes the coefficients corresponding to contaminated channels (``artifacts coefficients''), $\tilde{a}$ denotes the coefficients corresponding to the EMG channels, $i$ and $j$ are the number of uncontaminated channels and contaminated channels, respectively. Given that the ICs were normalized to unit variance, the coefficients contained in a given column of the mixing matrix (equation 5) can be interpreted as relative loads by which this IC contributed to the mixed signals. Therefore, in the artifact IC columns, large coefficient values are usually associated with the channels that were contaminated by artifacts, whereas uncontaminated channels would have relatively smaller coefficients. Satisfying this inequality (10) meant that the corresponding artifacts were detected in these artifact ICs.
After running ERASE 
on the combined simulated EEG data sets, all the artifact ICs, which were decided by the position of the maximal absolute values in the corresponding mixing matrix rows representing the EMG channels, were found. A false positive was formally defined as:
\begin{eqnarray}\label{10}
\fl \sum(|a_{i}^{*}|+...+|a_{i+j}^{*}|)/j - \sum(|a_{1}|+...+|a_{i-1}|+...+|a_{i+j+1}|+...+|a_{32}|)/(32-j)\nonumber\\> \max(|\tilde{a}_{33}|,...,|\tilde{a}_{36}|)* 5\%
\end{eqnarray}
 where 
 $\max(|\tilde{a}_{33}|,...,|\tilde{a}_{36}|)$ is defined as the maximal artifacts coefficient. If this inequality was satisfied in any artifact ICs, i.e. contaminated channels coefficients exceeded those for the uncontaminated channels by 5\% of the maximal artifacts coefficient (denoted as the threshold in Fig. 1), a false positive event occurred. The ratio of those false positive events in 200 simulated data sets was the false positive rate.

\textbf{Sensitivity:}
A sensitivity was designed to assess if contaminant EMG artifacts $N_{t}$ were accurately ``pushed'' into the artifact ICs, when contaminant EMG artifacts, $N_{t}$, were \emph{dependent on} reference EMG artifacts $n^{\ast}_{\tau}$. Sensitivity was defined as the ratio of events in which the contaminant EMG artifacts $N_{t}$ were detected in the artifact ICs after running ERASE. The simulated EMG which served as both contaminant EMG artifacts and reference EMG artifacts here, were used to contaminate the simulated EEG data as described above and also acted as separate channels. These combined simulated EEG data sets were subjected to ICA. Referring to the artifact ICs columns, the sensitivity in our study was calculated by the ratio of events in which the inequality (10) was satisfied in all the artifact ICs columns in corresponding 200 simulated data sets.



\subsection{Validation with real EEG}
The ability of ERASE to automatically reject real EMG artifacts from real EEG was assessed as follows.

\subsubsection{Experiments}

This study was approved by the Institutional Review Boards of the University of California, Irvine. Healthy subjects with no history of neurological conditions were recruited for this study. Subjects were fitted with a 64-channel EEG cap (ActiCap, Brain Products, Gilching, Germany) and asked to perform repetitive fist clenching and unclenching of the dominant hand while their EEG signals were acquired by two, linked NeXus-32 systems (Mind Media, Herten, Netherlands). EMG was recorded from the bilateral frontalis, left temporalis to masseter, right temporalis to masseter and bilateral trapezius using a MP150 system (BIOPAC, Goleta CA), respectively. The subjects were asked to sit in front of a computer screen, which prompted them to alternate between idling (for 5 s) or hand fist-clenching (for 2 s). This was repeated for a total of 10 trials over a 100s-long session. At least 2 sessions were performed by each subject. The EEG and EMG data were recorded at 2048 Hz and 4000 Hz sampling rates, respectively.

\subsubsection{Experimental data processing}

For the experimental EEG, both real and simulated EMG artifacts acted as separate channels and were not mixed to any EEG channels. The combined EEG/EMG data were bandpass filtered from 3 to 100 Hz (3$^{\text{rd}}$ order, forward-backward filter with no phase distortion). Note that the 100 Hz upper cutoff was chosen since the amplifiers attached to MP150 have a 100 Hz low-pass filter in hardware. Each trial, comprised 1-s idle time followed by 2-s movement, was identified and extracted from the combined EEG. Due to the non-stationarity of EEG, ICA decomposition was just applied to concatenated EEG trial datasets for each session (each run). The FastICA version in the EEGLAB toolbox \cite{15} was used to run ICA on the EEG trials data for all the subjects in the three conditions (simulated EMG, real EMG, and conventional ICA conditions). The artifact ICs were rejected as above. 
Short-time Fourier transform was applied to EEG trials and the signal power in different frequency bands ($\mu$ band: 8 to 12 Hz, high frequency (HF) band: 40 to 100 Hz) was compared across all ICA conditions.
All the data were z-scored after the time-frequency decomposition, which were separately normalized to the statistics of the EEG during the idling epochs.

The z-scored power of the $\mu$ and high-frequency bands during idle and movement was statistically compared for all the EEG channels using a Wilcoxon rank-sum test. The z-scored power of $\mu$/high-frequency band for each channel was then topographically mapped. For channels where there was no significant difference between idle and movement, these values were nulled (set to 0).

 Given that the absolute power at any frequency bands was reduced after artifact ICs were rejected from the original signal, we used the z-scored power of high-frequency band to calculate the decrease of high frequency after removal of EMG artifacts for each subject. The percent reduction was defined as below:
\begin{eqnarray}\label{11}
PD=\frac{|\sum^{C}_{c=1}\sum^{N}_{i=1}P^{b}_{z}(X_{i}^{c})-\sum^{C}_{c=1}\sum^{N}_{i=1}P^{a}_{z}(X_{i}^{c})|}{\sum^{C}_{c=1}\sum^{N}_{i=1}P^{b}_{z}(X_{i}^{c})}\times 100\%
\end{eqnarray}
where $P^{b}_{z}$ is the z-scored power of high frequency in baseline, $P^{a}_{z}$ is the z-scored power of high frequency after removal of EMG artifacts, $X$ are the EEG trials data, $i$ is the $i^{th}$ trial, $N$ is the total number of available trials for each subject, $c$ is the $c^{th}$ channel, $C$ is the total number of EEG channels. The rationale for this approach is that the high-frequency signal is dominated by EMG in EEG and any reduction in the high-frequency band power was considered as the reduction in EMG.
The Wilcoxon rank-sum test was also employed to examine the difference of the $\mu$ band power in the C3/C4 channel and the high-frequency band power in all of the channels during movement between all combinations of the 4 conditions (baseline, after ERASE with simulated EMG, after ERASE with real EMG, and after conventional ICA).
\begin{figure}
  \centering
  \vspace{-0.1in}
  \includegraphics[width=\textwidth]{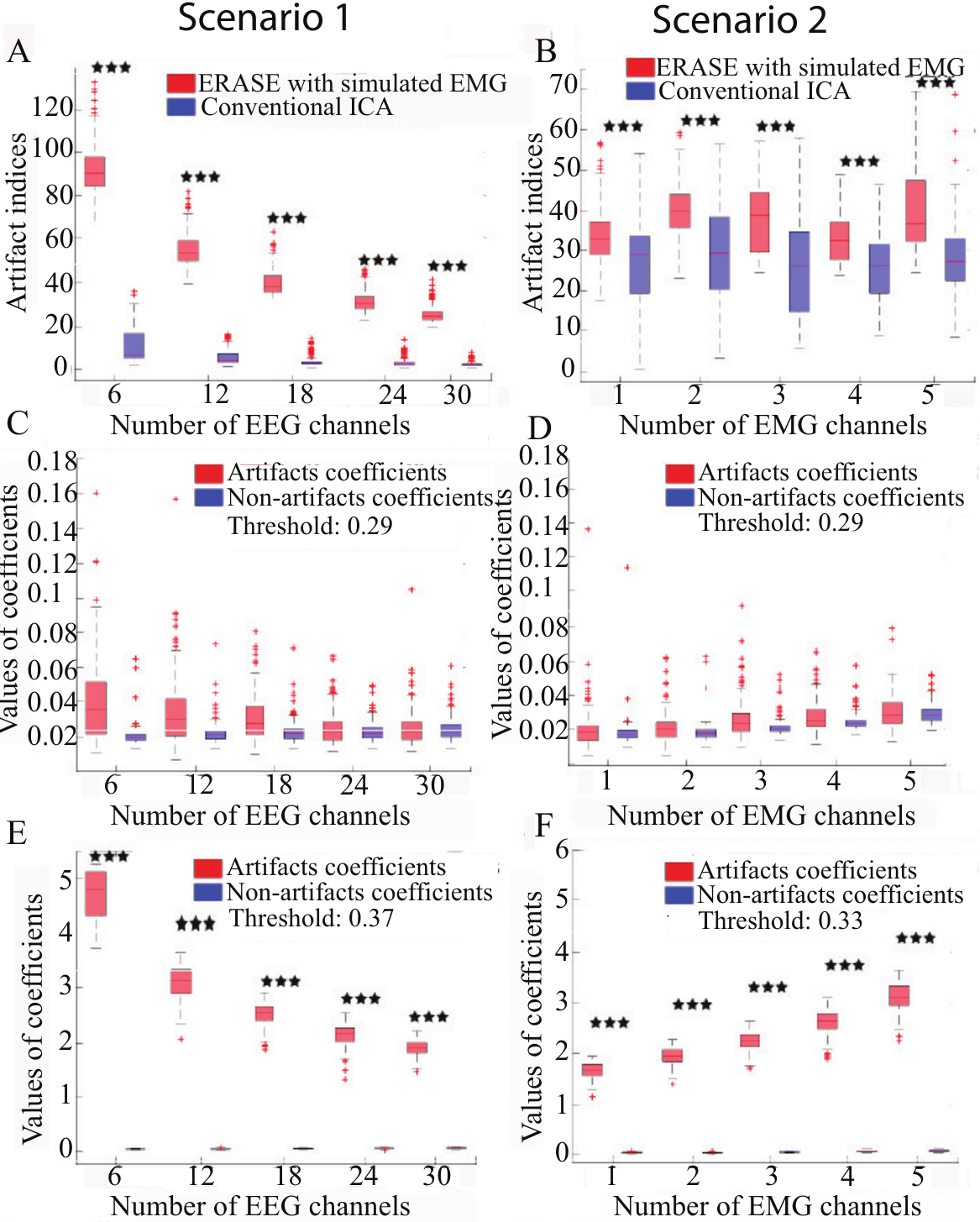} \vspace{-0.3in}
  \caption{Effectiveness, false positive, and sensitivity of ERASE on the simulated data. \textbf{A, B.} Artifact indices, i.e., effectiveness in Scenario 1, 2 respectively. \textbf{C, D.} False positive for Scenario 1 and 2, respectively. \textbf{E, F.} Sensitivity results for Scenario 1 and 2, respectively. Scenario 1 was considered as the fact that the number of contaminated EEG was changed for the performance assessment of ERASE. Here, the number of contaminated EEG channels were chosen from 2 to 10 for each simulated EMG artifacts (three EMG artifacts were used here, so numbers are 6 to 30 in figures). We increased the number of added EMG channels in Scenario 2 for the performance assessment of ERASE. The number of added EMG channel was chosen from 1 to 5. Asterisks indicate a significant difference between two data sets (Wilcoxon Rank-Sum Test), and the significance level=***p$<$0.001.}\label{fig.1}
  \vspace{-0.2in}
\end{figure}
\section{RESULTS}
\subsection{Simulation verification}

\begin{itemize}
  \item[a.] \textbf{Effectiveness: }
  The artifact indices for ERASE and conventional ICA were summarized in Fig. 1 A and B. Across all parameters, the artifact indices calculated by equation 8 in ERASE were significantly larger than those in the conventional ICA model (Wilcoxon rank-sum test, P$<$0.001). This indicates that ERASE has better effectiveness in removing the EMG artifacts compared to the conventional ICA model.
  \item[b.] \textbf{False Positive Rate: }
  The artifacts and non-artifacts coefficients for the false positive were summarized in Fig. 1 C and D. Across all parameters, there was a 0\% rate of false positive, as defined by Inequality 10 (details regarding thresholds are in Supplementary Table 1). Furthermore, the values of artifacts and non-artifacts coefficients were small and were not significantly different from one another. This indicates that signals independent of the reference artifacts were not erroneously ``pushed" into artifact ICs by ERASE.
  \item[c.] \textbf{Sensitivity: }
  The artifacts and non-artifacts coefficients for the sensitivity were summarized in Fig. 1 E and F. Across all parameters, there was a 100\% rate of sensitivity, as defined by Equation 10 (details about threshold were shown in Supplementary Table 2). Furthermore, the values of artifacts and non-artifacts coefficients were significantly different from one another (Wilcoxon rank-sum test, P$<$0.001) and the values of non-artifacts coefficients were always significantly lower. This indicates that contaminant EMG artifacts can be accurately identified by ERASE.
\end{itemize}

\begin{figure}
  \centering
  \vspace{-0.2in}
  \includegraphics[width=\textwidth]{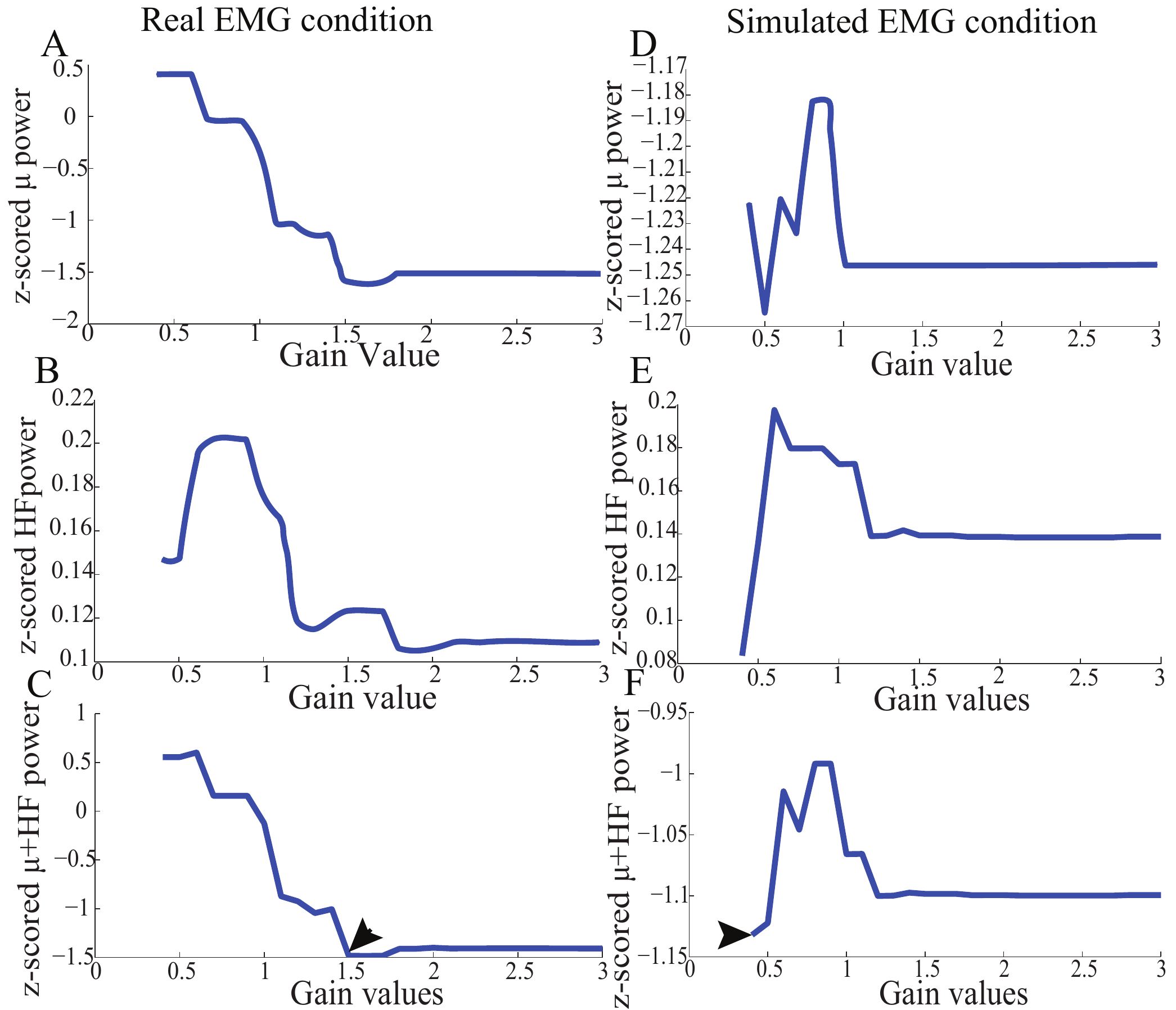} \vspace{-0.3in}
  \caption{Representative z-scored $\mu$ and high-frequency band power from Subject 1 after running ICA (as in Section 2.3.2), using simulated EMG and real EMG. \textbf{A.} Average of z-scored $\mu$ power in the C3 channel in the real EMG condition. \textbf{B.} Average of z-scored high-frequency band power for all the EEG channels in the real EMG condition. \textbf{C.} Values added by z-scored high-frequency band and $\mu$ power in the real EMG condition. \textbf{D.} Average of z-scored $\mu$ power in the C3 channel in the simulated EMG condition. \textbf{E.} Average of z-scored high-frequency band power for all the EEG channels in the simulated EMG condition. \textbf{F.} Values added by z-scored high-frequency band and $\mu$ power in the simulated EMG condition. After running ICA on the combined EEG/EMG data, the artifact ICs were rejected based on the criteria depicted above for each threshold. After the joint consideration, the final threshold was 1.5 for real EMG condition, and 0.4 for simulated EMG, which were denoted by arrows.}\label{fig.2}
  \vspace{-0.2in}
\end{figure}

\begin{table}[htb]
\tiny
\centering
\caption{Change of HF band power in different conditions for each subject ( mean $\pm$ standard deviation (S.D))}\label{table1}
\begin{threeparttable}
\begin{tabular}{p{0.18\columnwidth}p{0.08\columnwidth}p{0.08\columnwidth}p{0.08\columnwidth}p{0.08\columnwidth}p{0.09\columnwidth}p{0.08\columnwidth}p{0.08\columnwidth}p{0.08\columnwidth}}
  \hline
 \textbf{Subject} &S1 & S2 & S3 & S4$^{*}$ & S5 & S6 & S7 & S8$^{*}$ \\

    \hline
     \multicolumn{8}{l}{\textbf{Baseline}}\\
    Z-scored HF band& 0.35$\pm$0.26 & 0.18$\pm$0.14 & 0.35$\pm$0.24& 0.24$\pm$0.163& 0.04$\pm$0.04& 0.18$\pm$0.13& 0.50$\pm$0.19& 0.71$\pm$0.25\\
    Z-scored $\mu$ band & -0.86$\pm$0.58 & -0.13$\pm$0.04& -0.01$\pm$0.07 & -0.91$\pm$0.58 & -0.48$\pm$0.18& -0.97$\pm$0.72& -0.57$\pm$0.49& 0.04$\pm$0.32\\
    \hline
     \multicolumn{8}{l}{\textbf{After ERASE with real EMG}}\\
    Z-scored HF band& 0.12$\pm$0.06 & 0.06$\pm$0.07 & 0.03$\pm$0.02& 0.11$\pm$0.06 & 0.40$\times$10$^{-4}$ $\pm$0.01& 0.02$\pm$0.30 $\times$10$^{-2}$& 0.10$\pm$0.08& 0.31$\pm$0.03\\
     Z-scored $\mu$ band &-0.78$\pm$0.38 & -0.43$\pm$0.18& -0.40$\pm$0.25 & -1.03$\pm$0.39& -0.48$\pm$0.26& -1.03$\pm$0.48& -0.67$\pm$0.18& -0.82$\pm$0.56\\
        Reduction percentage (\%)& 64.42 & 67 & 90.96 & 54.19 & 99 & 91.4 & 79.94 & 55.57 \\
    \hline
     \multicolumn{8}{l}{\textbf{After ERASE with simulated EMG}}\\
    Z-scored HF band& 0.12$\pm$0.16 & 0.08$\pm$0.08& 0.06$\pm$0.06& 0.12$\pm$0.10& 0.50$\times$10$^{-4}$ $\pm$0.50$\times$10$^{-2}$& 0.03$\pm$0.08& 0.15$\pm$0.08& 0.31$\pm$0.12\\
    Z-scored $\mu$ band & -0.92$\pm$0.44 & -0.31$\pm$0.24& -0.32$\pm$0.13& -0.90$\pm$0.49 & -0.33$\pm$0.34& -0.90$\pm$0.59& -0.50$\pm$0.12 & -0.84$\pm$0.35\\
    Reduction percentage (\%)& 65.77 & 56.16 & 83.62 & 50 & 99 & 83.08 & 69.79 & 56.45 \\
  \hline
   \multicolumn{8}{l}{\textbf{Conventional ICA}}\\
    Z-scored HF band& 0.17$\pm$0.14 & 0.09$\pm$0.07& 0.17$\pm$0.11 & 0.17$\pm$0.12 & 0.70$\times$10$^{-2}$ $\pm$0.8$\times$10$^{-2}$& 0.08$\pm$0.01& 0.19$\pm$0.09& 0.41$\pm$0.13\\
     Z-scored $\mu$ band & -0.57$\pm$0.72 & -0.21$\pm$0.35 & 0.10$\pm$0.12& -0.58$\pm$0.35& -0.39$\pm$0.29 & -0.58$\pm$0.72 & -0.50$\pm$0.17 & -0.70$\pm$0.30\\
     Reduction percentage (\%)& 51.8 & 52.96 & 52.54 & 26.76 & 82.06 & 56.99 & 61.92 & 41.99 \\
  \hline
\end{tabular}
\begin{tablenotes}
        \small
 \item HF: high frequency.
 \item z-scored HF band: the average z-scored power during movement over all the available sessions.
 \item Reduction percentage: the difference between high-frequency band power in baseline and after ERASE or conventional ICA was divided by high-frequency band power in the baseline.
 \item Twenty trials were employed for calculation for Subject 4 and 8 (denoted by asterisks). Ten trials were used for the remaining subjects.
      \end{tablenotes}
    \end{threeparttable}
    \vspace{-0.2in}
\end{table}
\subsection{Experimental Verification}
A total of 8 subjects gave their informed consent to participate in the study. All subjects were healthy male volunteers between 20 and 50 years old. A total of 12 sessions (120 trials) were performed across all subjects.

 \begin{figure}[!h]
  \centering
  \vspace{-0.1in}
  \includegraphics[width=\textwidth]{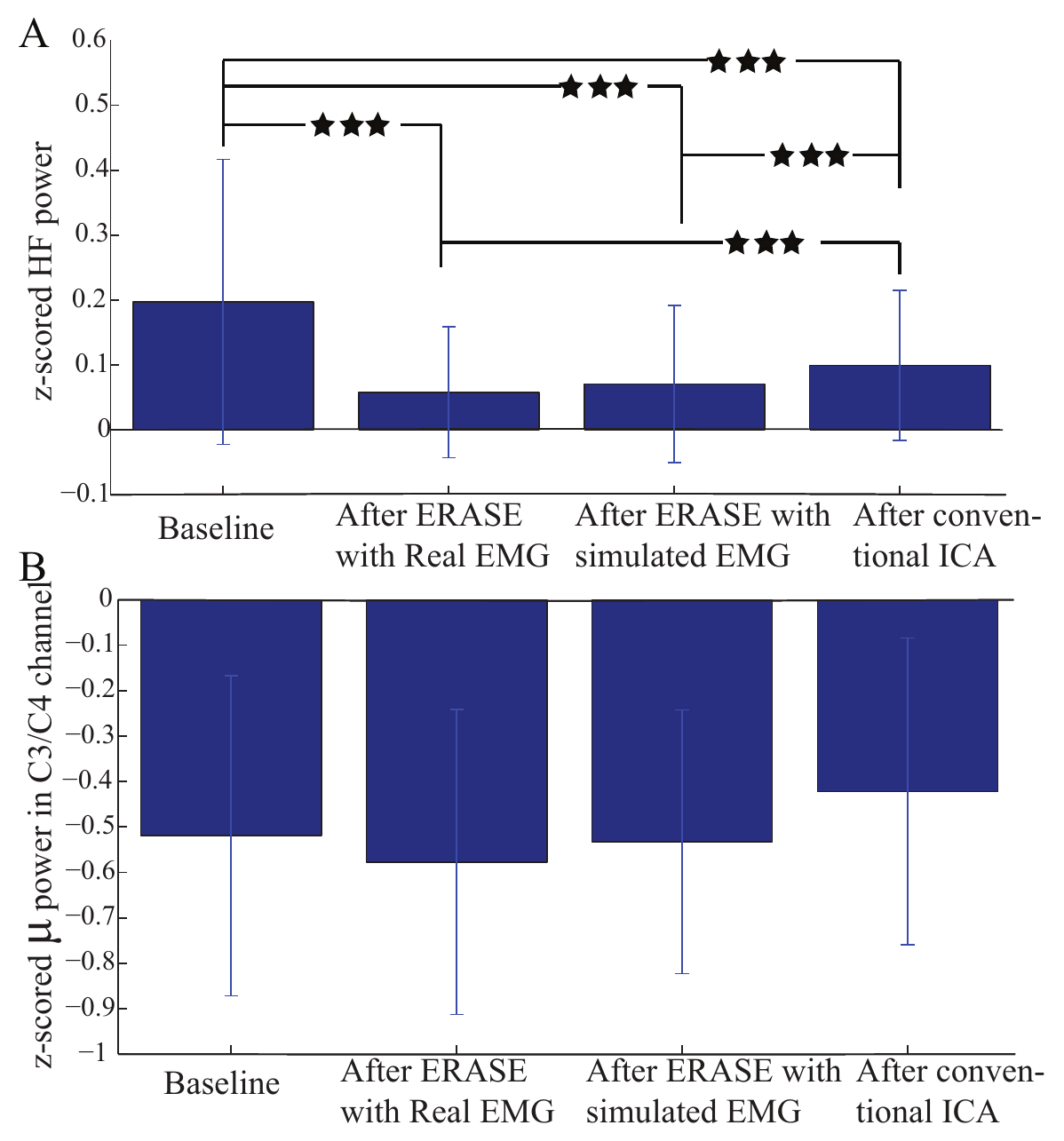} \vspace{-0.3in}
  \caption{Comparison of z-scored power of $\mu$/high-frequency band during movement in different conditions. \textbf{A.} z-scored power of high-frequency band during the movement for all the EEG channels under different conditions. \textbf{B.} z-scored power of $\mu$ band during movement in the C3/C4 channel under different conditions. Data were from all the subjects with a total of 120 trials. ***: significant differences between the two datasets (p$<$0.001).}\label{fig.3}
\vspace{-0.2in}
\end{figure}


An example of selecting the proper threshold for ERASE was shown in Fig. 2. Table 1 summarized the effect of ERASE on the $\mu$-band and high-frequency band across all subjects. Overall, the high-frequency band power, typically dominated by EMG artifacts, was reduced by 75.31\% using ERASE with real EMG, by 63.46\% through ERASE with simulated EMG and only by 48.88\% with the conventional ICA approach (Table 1). At the group level (Fig. 3 A), the z-scored power of the high-frequency band during movement was significantly reduced after running ERASE. Furthermore, the high-frequency band power in the real EMG and simulated EMG conditions were both significantly lower than that in the conventional ICA (Fig. 3 A). However, there was no difference between running ERASE with either the real EMG or simulated EMG. At the group level, there was no change in the $\mu$ band (only for C3/C4 channel) among the four conditions based on all trials (Fig. 3 B). This indicates that the expected $\mu$-band desynchronization phenomenon during the fist-clenching task was not adversely affected by ICA.
\begin{figure}[!h]
  \centering
  \vspace{-0.1in}
  \includegraphics[width=\textwidth]{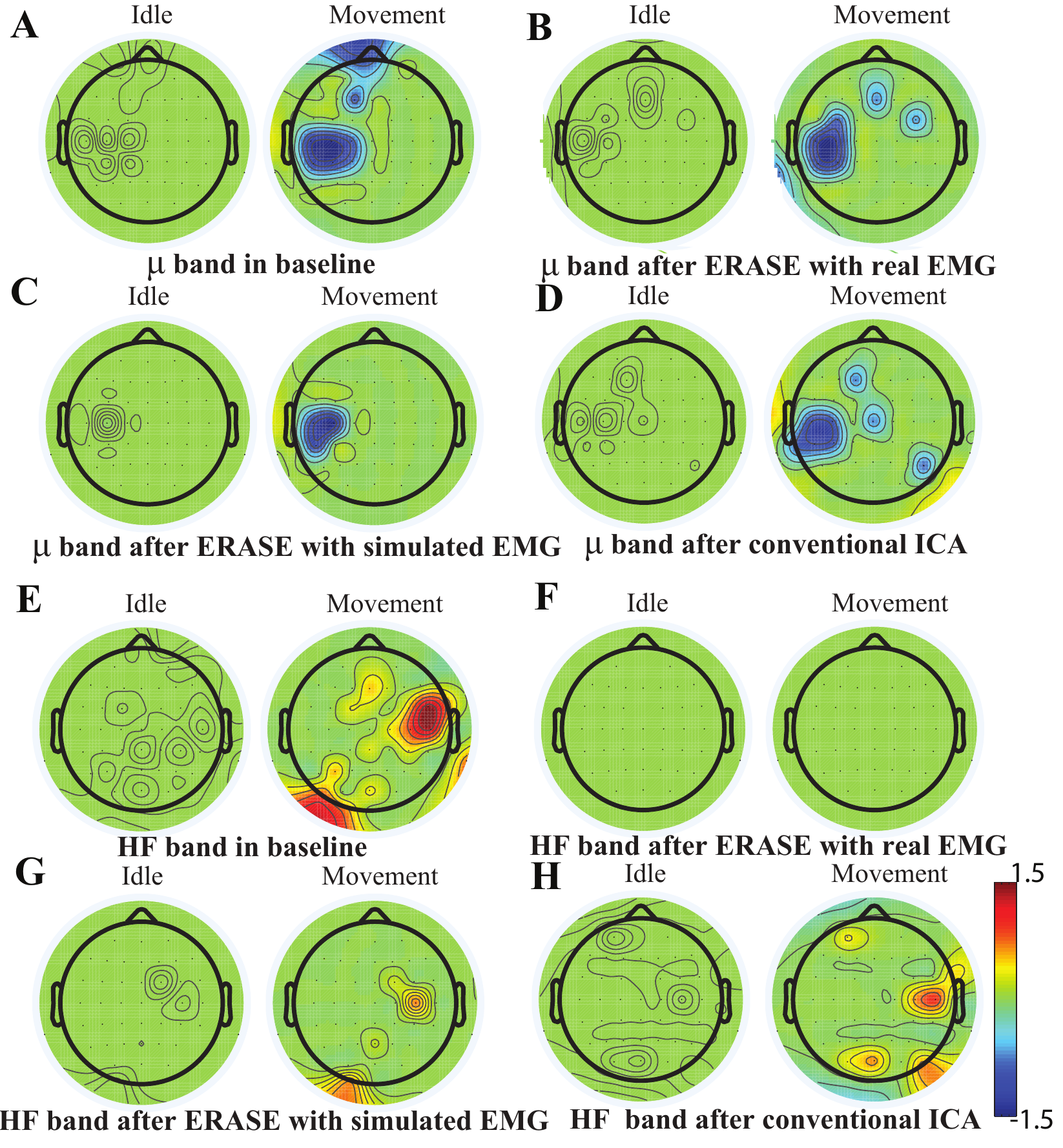} \vspace{-0.3in}
  \caption{Brain topography maps for Subject 1 displaying the z-scored power of $\mu$ band and high-frequency band in different conditions (baseline, after ERASE with real EMG and with simulated EMG and after conventional ICA, \textbf{A-D} for $\mu$ band and \textbf{E-H} for high-frequency band) on the Subject 1. Channels whose z-scored power of $\mu$/high-frequency band were not significantly different between idle and movement states (Wilcoxon rank-sum test) were nulled (values were set to zero). The significance threshold was P$<$0.01 for $\mu$ band and P$<$0.05 for high-frequency band. }\label{fig.3}
\vspace{-0.2in}
\end{figure}

 A representative example of significant changes in z-scored power of $\mu$ and high-frequency bands during idle and movement is shown in Fig. 4. The z-scored power of high-frequency band during movement was reduced after running ERASE, and the power in both ERASE conditions was smaller than that in the conventional ICA (Fig. 4 E-H, Supplementary Fig. 1-7 E-H. A representative example of time series also showed the same findings (Supplementary Fig. 9)). In Fig. 4 A-D, the $\mu$ desynchronization during right hand movement was well preserved after running ICA in all the conditions and was localized to the C3 channel and surrounding electrodes. This localized $\mu$ desynchronization was also present around the C3 channel for Subjects 3 to 7 (Supplementary Fig. 1-5 A-D) and around the C4 channels for Subjects 2 and 8 (Supplementary Fig. 6-7 A-D). The channels exhibiting $\mu$ desynchronization were always contralateral to the hand movement except Subjects 2 and 8 (Supplementary Figs.1 and 7). Combined with the findings above, this indicates that ERASE did not disturb the spatial distribution of the expected brain features underlying the motor task of interest.


\section{Discussion}
Here, we proposed a modified ICA model that combined reference EMG artifacts with EEG data to facilitate an enhanced automated removal of EMG artifacts. We tested and validated this method using both simulated and actual EEG during hand movement. We found that it had high sensitivity at detecting EMG artifacts and an extremely low false positive rate (Fig. 1 C-F, Supplementary Table 1 and 2). With simulated data, ERASE effectively removed a large proportion of the EMG artifacts (Fig. 1 A-B). It also removed EMG artifacts in real EEG recordings, while preserving the expected $\mu$ desynchronization associated with movement (Fig. 4 and Supplementary Figs. 1-7). This may also indicate that our approach can remove any potential confounding overlap between EMG and EEG and thereby improves the confidence that low-frequency brain features extracted from ERASE are mostly EMG-free (which cannot be achieved by a simple low pass filter).

These results suggest that using reference EMG artifacts can force ICA to ``learn'' and detect the EMG artifacts by forcing the contaminant EMG within EEG into a minimal number of ICs. We established an operational definition (rejection criteria) for identifying EMG artifacts components to enable automated component rejection and thereby minimizing user bias. Compared to conventional ICA, results showed that ERASE removed on average 26\% more EMG artifacts from EEG data than conventional ICA (Fig. 3 and Table 1), which indicated that ERASE improved the ICA algorithm. Although this approach may require slightly more preparation time to record real EMG, it is still possible to use it in situations where real EMG recordings were either not possible or not available by substituting it with simulated EMG (Fig. 3). The advantages and novelty of this approach are discussed in further detail below.

First, ERASE directly introduces reference EMG artifacts into the ICA model as prior knowledge to more accurately maximize the separation between EMG and EEG ICs as well as to minimize the computational complexity of removing EMG with respect to existing cICA approaches. On the other hand, previously reported forms of reference ICA, such as spatially cICA \citep{a48,a49}, involves first performing an initial run of ICA on a previous EEG data segment, followed by manually selecting ICs believed to represent EMG sources as EMG reference. Subsequently, ICA is run iteratively on current EEG data to find all the ICs which have a strong correlation with pre-defined EMG reference. By comparison, ERASE real EMG (where available) or simulated EMG as the reference. Since such references are expected to provide a more reliable representation of the ground truth for EMG sources, ERASE is is likely more reliable and systematic than other forms of ICA. In addition, ERASE utilizes properties of mixing matrix for identification of artifact ICs to avoid the complicated computation of optimization problem, which is employed in temporally cICA \cite{a11,13,a47}. 
When combined with the use of a simulated or simultaneously recorded real EMG used in ERASE, it is not necessary to perform iterative run for the ICA process. Our results (Figs. 3 and 4 and Table 1) showed that ERASE can remove most of EMG artifacts and preserve expected brain features just after a single ICA run.


Unlike previously reported versions of ICA, ERASE does not require manual intervention. More specifically, ICA is typically run directly on EEG, and the operator manually rejects the artifact ICs. The effectiveness of this step depends substantially on the experience of researchers and may be biased due to subjectivity. Hence, ERASE can minimize the biases of researchers and improve the efficiency of artifacts rejection. As mentioned in the introduction, automated rejection is not necessarily unique to ERASE \cite{a14,a12,a52,7,8,a44,a54,51,52}, given that other methods, such as cICA can also involve automatic IC rejection when prior knowledge of EMG signals is available \cite{a48,a49,a1}.
However, one unique aspect of ERASE compared to these prior reports is that rejection criteria are based on physiological features of both EEG and EMG for automated EMG artifacts rejection procedure (Section 2.1.2), which makes ERASE more focused on preserving relevant EEG phenomenon.

ERASE was validated with both simulated and behavioral EEG data, whereas the physiological information and properties of EEG are completely overlooked in other EMG artifact removal studies using ICA, cICA, or other BSS methods. Researchers typically employ simulated or synthetic EEG data to validate corresponding artifacts removal algorithms. Based on various metrics (discussed in the paragraph below), all of these algorithms have declared themselves effective at removing EMG artifacts, but have not answered a critical question as to whether the information encoded in EEG is retained after artifact removal. Here, we show that EEG $\mu$ band modulation that typically underlies hand movement is preserved or even enhanced after ERASE. Combined with the reduction of EMG elsewhere in the EEG, these findings unequivocally demonstrate that meaningful EEG is retained. Such a demonstration has been generally absent from the validation of other ICA approaches.

 It should be noted that ICA was selected as the basis of ERASE due to the fact that ICA has typically been shown to have superior performance to most other artifact removal methods. For example, CCA, a popular algorithm for artifact removal, does not outperform ICA at removing EMG artifacts from EEG \cite{a1,a15,a37,a38} as well as at removing ECG and EOG artifacts \cite{a39,a40,a41,a42,a43,a44}. Some BSS methods and source decomposition methods have been combined for removal of EMG artifacts (e.g. EEMD-CCA \cite{55,56,57}, as well as EEMD-ICA \cite{58}). Both EMD and EEMD are single-channel techniques, so EEMD-CCA and EEMD-ICA are only tested on the fact that a few channels of EEG recording are involved. Since running ICA or CCA sometimes is time-consuming, EEMD-CCA or EEMD-ICA probably is typically less than ideal. 

 There are no uniformly accepted performance metrics for artifacts removal algorithm in practical EEG. In most studies \cite{59,29,a61}, the performance of EMG artifact removal is typically assessed, in part, by visual inspection. Although highly subjective, it may give an indication with respect to whether the algorithm has improved the quality of the EEG signal or has distorted one or more time intervals or frequency bands. This is also performed in our study and the results are showed in Supplementary Fig. 9. For simulated and synthetic EEG, there are many metrics employed for the assessment of the performance of artifact removal approaches, since the `ground truth' in such scenarios are explicitly known \cite{a15,a62,a49,a63,58,a64}. The most widely used metrics for performance assessment are listed in the literature \cite{a1} (e.g. the relative root mean squared error (RRMSE), the signal to artifact ratio (SAR), etc.). In our work, we used standard statistical metrics (false positive, sensitivity, effectiveness) to validate our approach in simulated EEG. Meanwhile, time-series validation results are shown in Supplementary Figs. 10-12 in the Appendix. It should also be emphasized that since researchers in each study used different data sets, a head-to-head performance comparison across various artifact removal approaches in EEG studies is difficult. Therefore, we further validated our novel approach by using an assessing both the elimination of EMG and the preservation of the EEG features underlying motor behaviors. We propose that such an evaluation is employed for validation of effectiveness of any new artifact rejection approaches in future studies, as it is otherwise not possible to know if the methodology aggressively removed EMG as well as erroneously eliminating the signals of interest.

The main limitations of ERASE are that it is still impossible to remove all EMG from EEG. The main reason originates from the assumptions of our theoretical model. Namely, our model requires that the reference EMG artifacts and contaminant EMG artifacts are completely dependent. However, contaminant EMG artifacts in EEG data cannot have complete dependence on the real EMG sources due to several reasons. For example, the signal propagation path between the true source and recording electrodes may significantly distort observed EMG. It is also not possible to include all the reference EMG artifacts which could be contributing to EEG, because some head and neck muscles are not easily recorded at the surface. In addition, another assumption in our theoretical derivation is that the reference EMG artifacts are independent of EEG. Also, the reference EMG electrodes are close to potential EEG sources (such as EMG reference electrodes located over the frontalis and temporalis muscles), so it is difficult to ensure that the reference EMG artifacts channels in this study are fully independent of EEG. Hence, the identified artifact ICs may still contain some contribution from EEG. However, EEG from these areas is not highly involved in motor tasks in our studies.

\section{CONCLUSION}
Here, we proposed a modified ICA model that can automatically remove EMG artifacts by combining reference EMG artifacts with EEG. This new approach can more effectively remove EMG artifacts from EEG while preserving the expected brain features underlying motor behavior. Also, the approach proposed in this work is automated, which minimizes experimenter bias and speeds up analysis. The utilization of the simulated EMG as the reference EMG source potentially extends the application of this approach. The EEG recovered by our approach can provide more confidence for further neuroscience analysis. Meanwhile, future work will focus on testing ERASE on EEG from patient populations, and adapting it for real-time applications, such as in the BCI system.

\section{Appendix}
Appendix: We showed the results of z-scored power of $\mu$ band (8 to 12 Hz) and high frequency band (40 to 100 Hz) in different conditions (baseline, after ERASE with real EMG and simulated EMG and after conventional ICA) for Subject 2-8 (Figs. 1-7), the 2D image of the electrode locations (Fig. 8), time series band-pass filtered with 40-100 Hz for one trial in different conditions (Fig. 9), time series validation results by using simulated EEG with simulated EMG (Fig. 10) and real EMG (Fig. 11), and performance comparison between ERASE and conventional ICA by a common metric. Also, Table 1 and 2 list the average thresholds for false positive and sensitivity test, respectively. Table 3 lists the average artifact indices for the simulated EMG and conventional ICA conditions.





%
\newcommand{\newblock}{}
\bibliographystyle{unsrt}
\bibliography{reference}

\section{Supplementary}
\begin{figure}[!h]
  \centering
  \includegraphics[width=\textwidth]{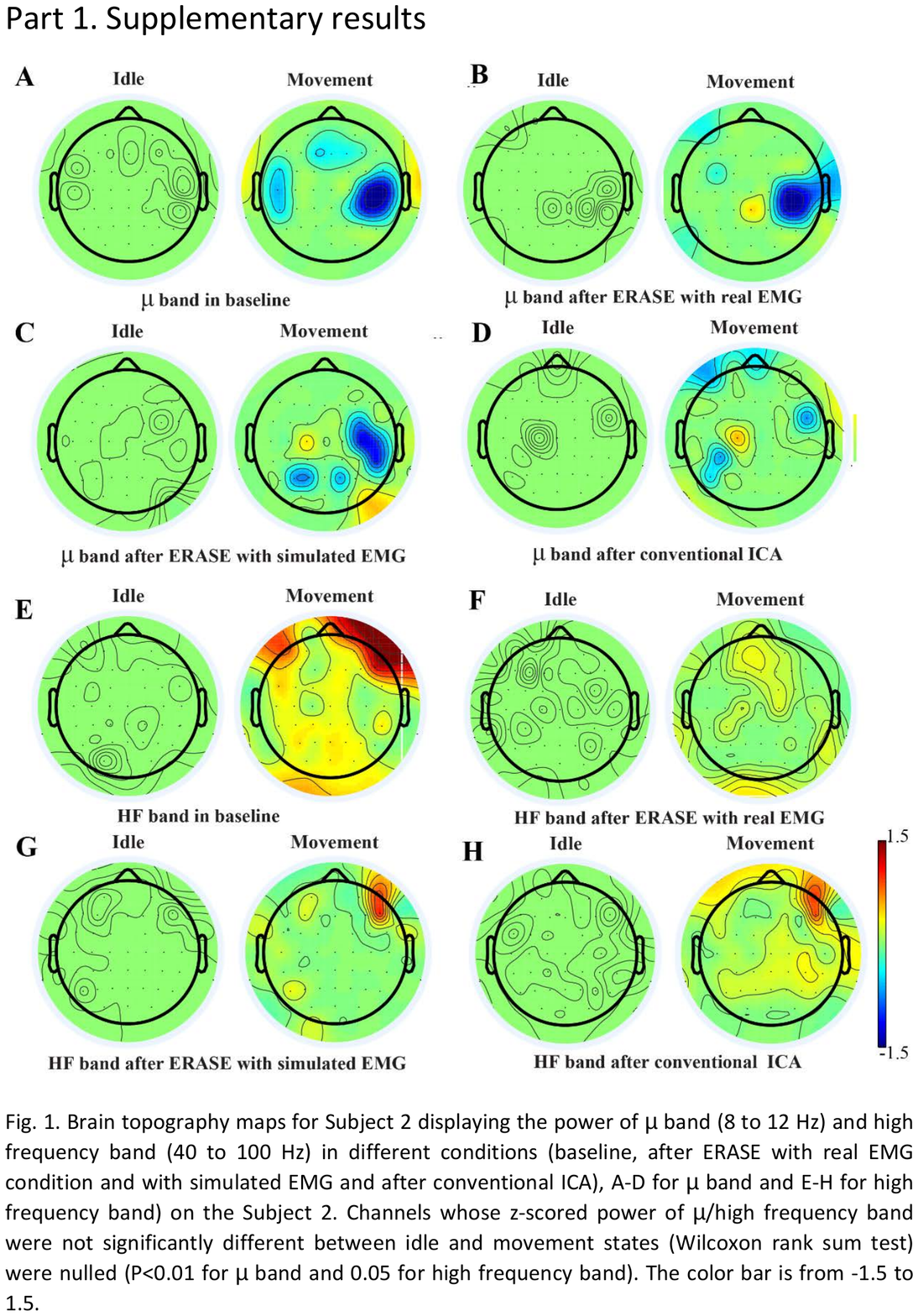} 
\end{figure}

\begin{figure}[!ht]
  \centering
  \includegraphics[width=\textwidth]{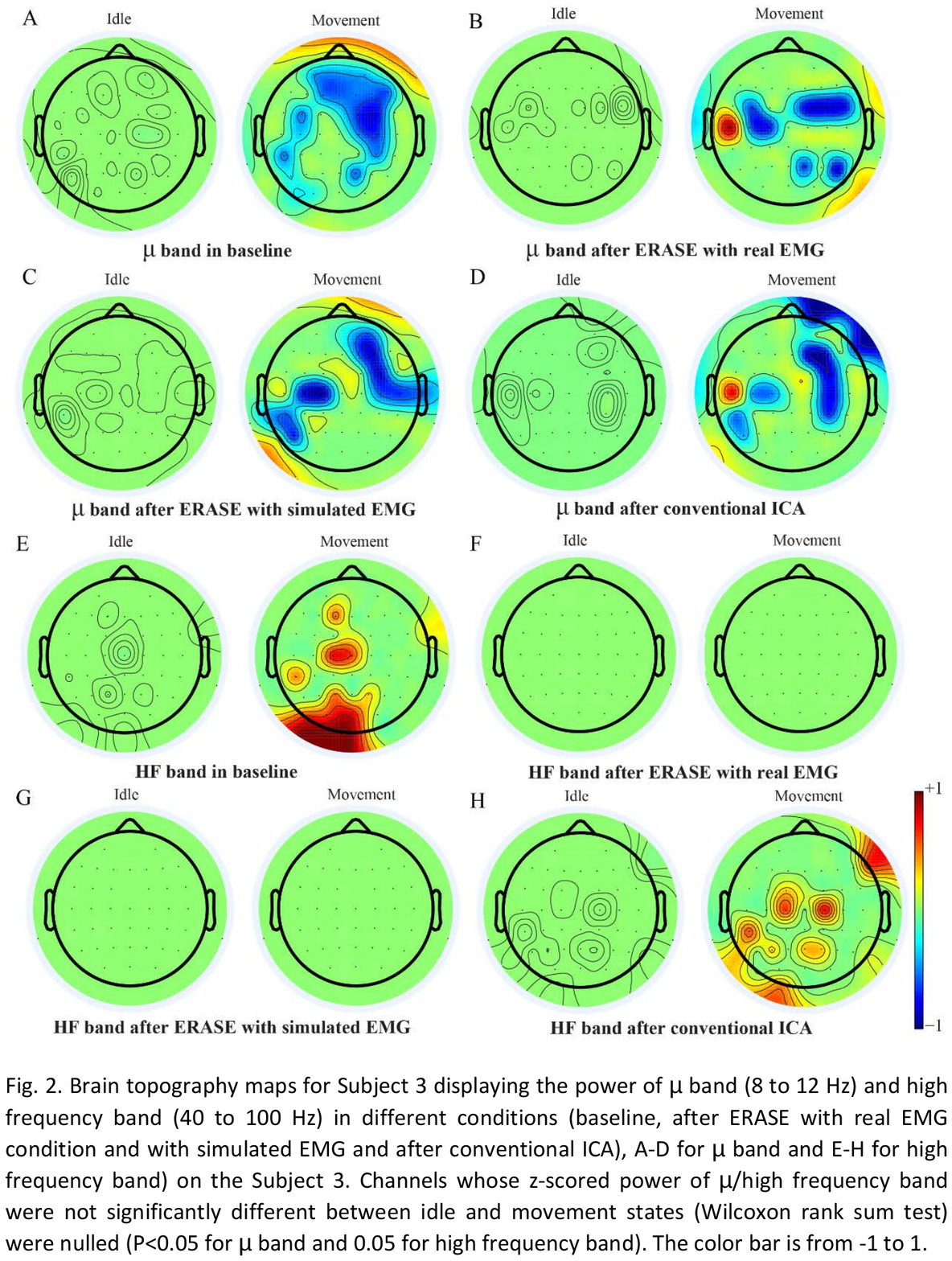} 
\end{figure}

\begin{figure}[!ht]
  \centering
  \includegraphics[width=\textwidth]{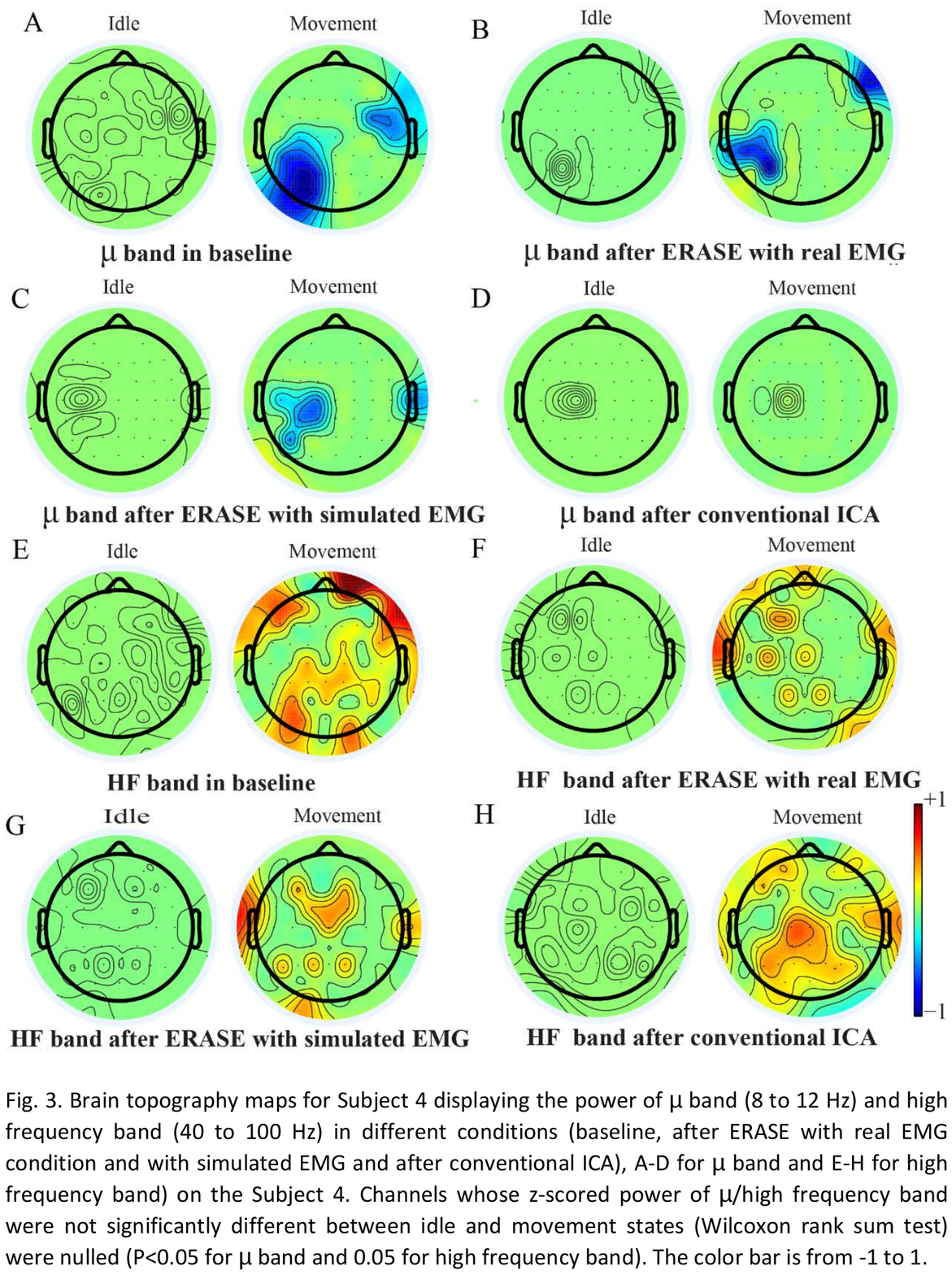} 
\end{figure}

\begin{figure}[!ht]
  \centering
  \includegraphics[width=\textwidth]{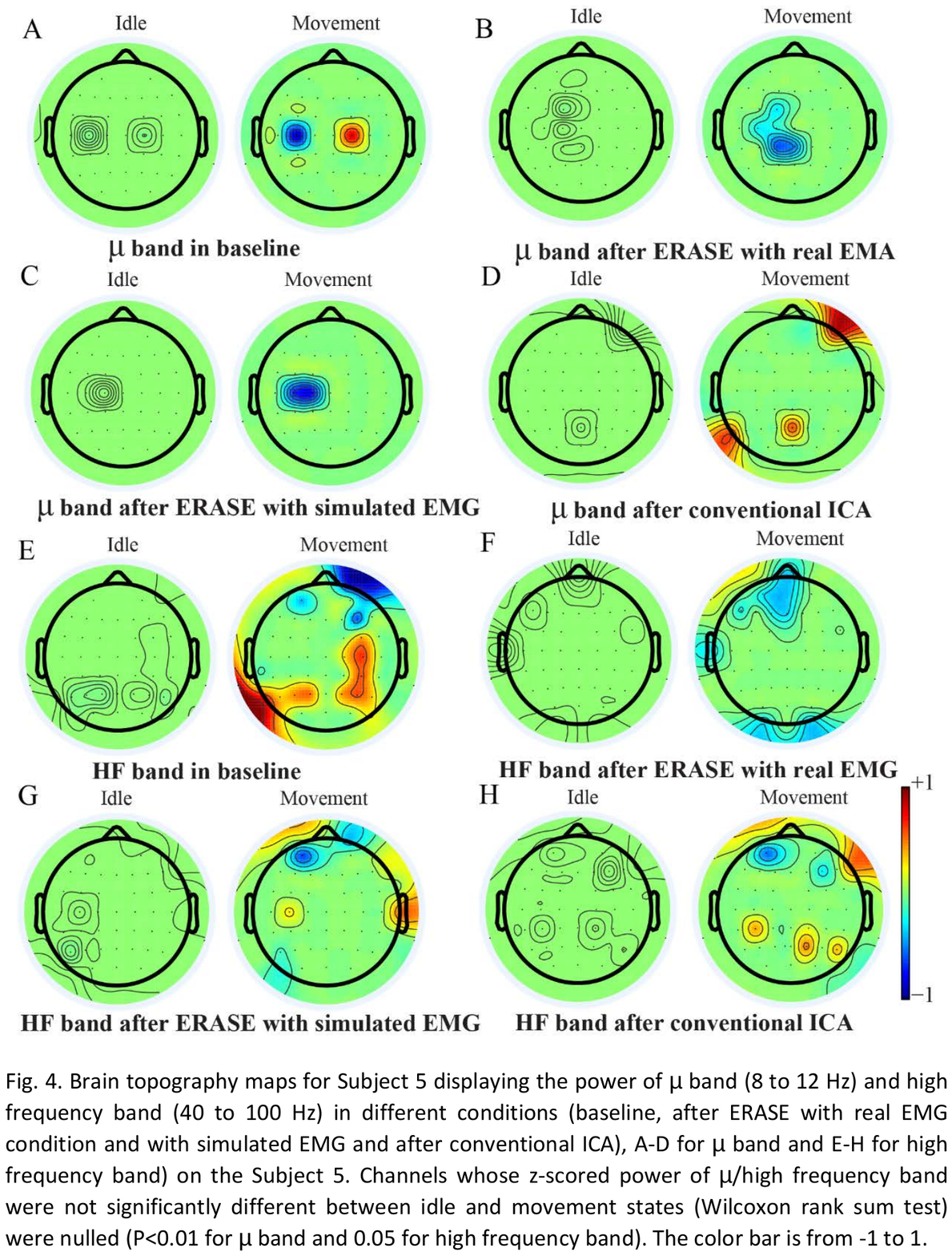} 
\end{figure}

\begin{figure}[!ht]
  \centering
  \includegraphics[width=\textwidth]{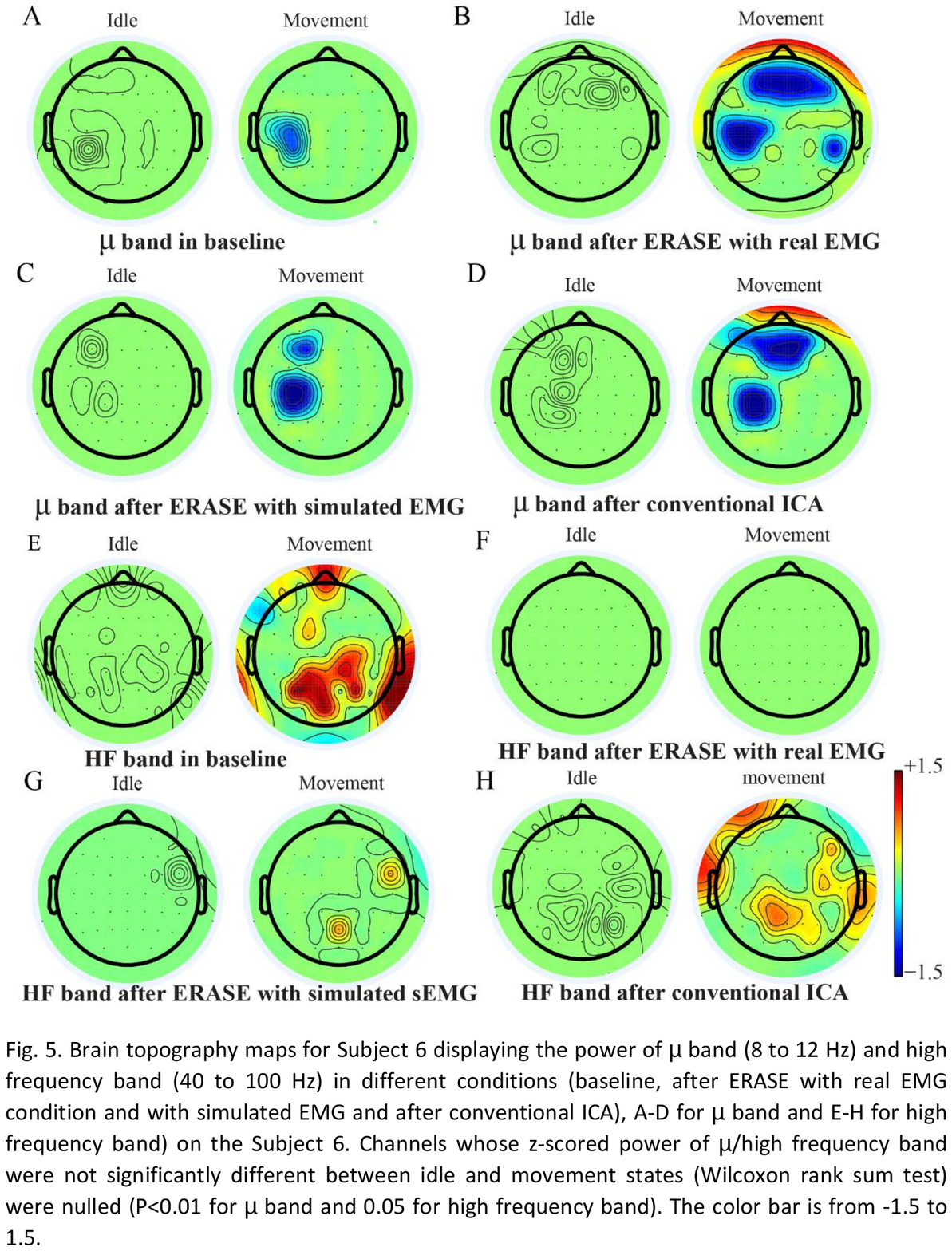} 
\end{figure}

\begin{figure}[!ht]
  \centering
  \includegraphics[width=\textwidth]{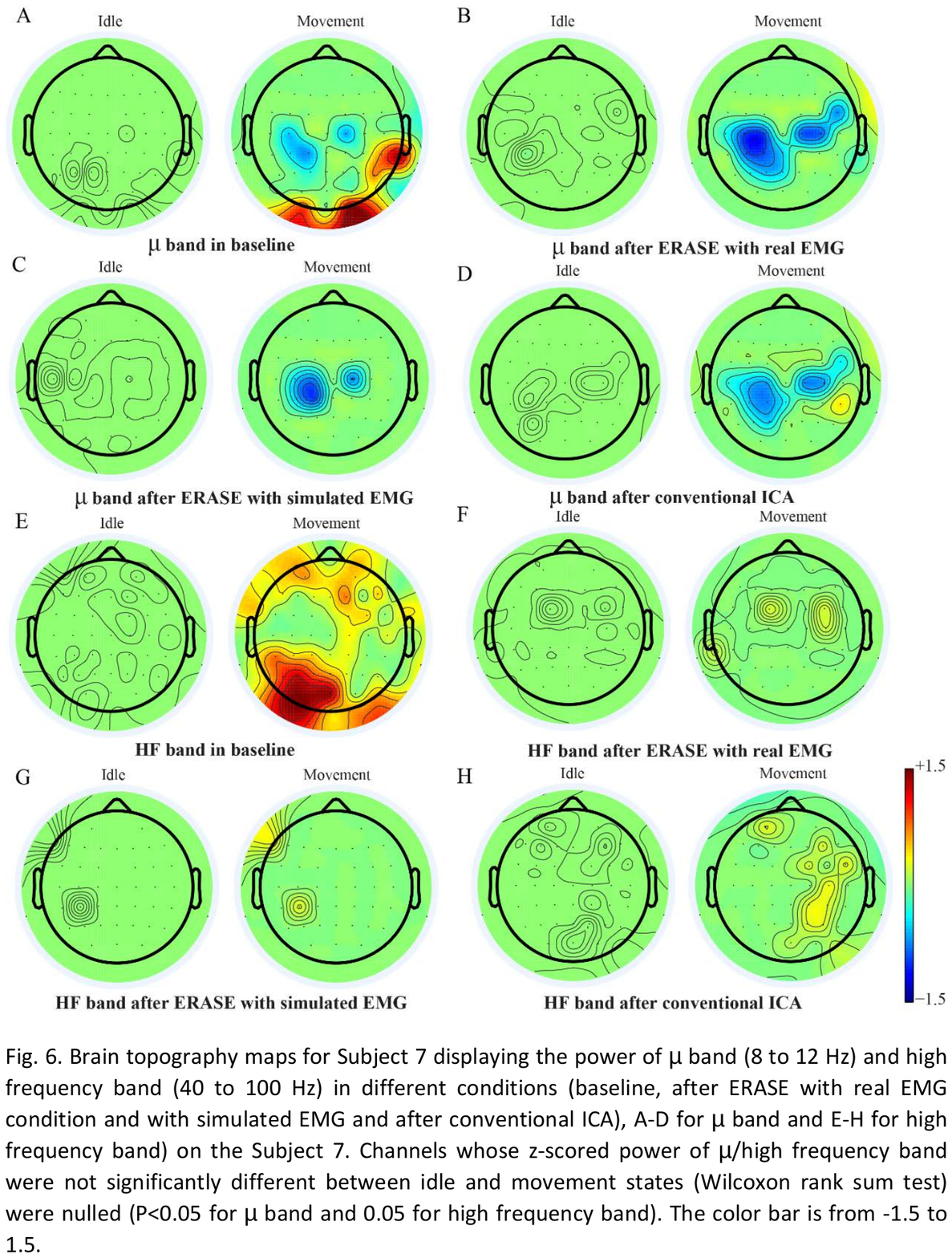} 
\end{figure}

\begin{figure}[!ht]
  \centering
  \includegraphics[width=\textwidth]{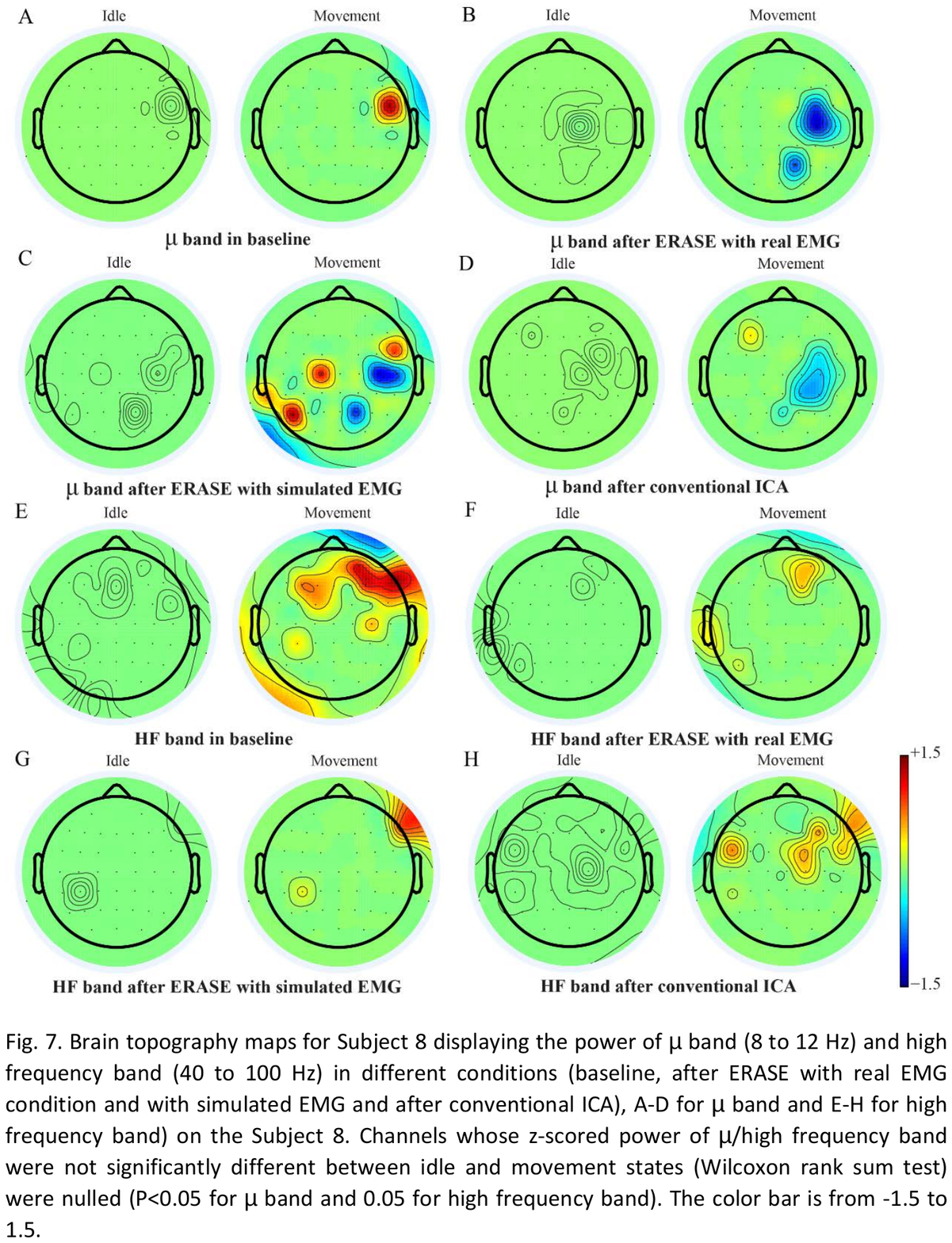} 
\end{figure}

\begin{figure}[!ht]
  \centering
  \includegraphics[width=\textwidth]{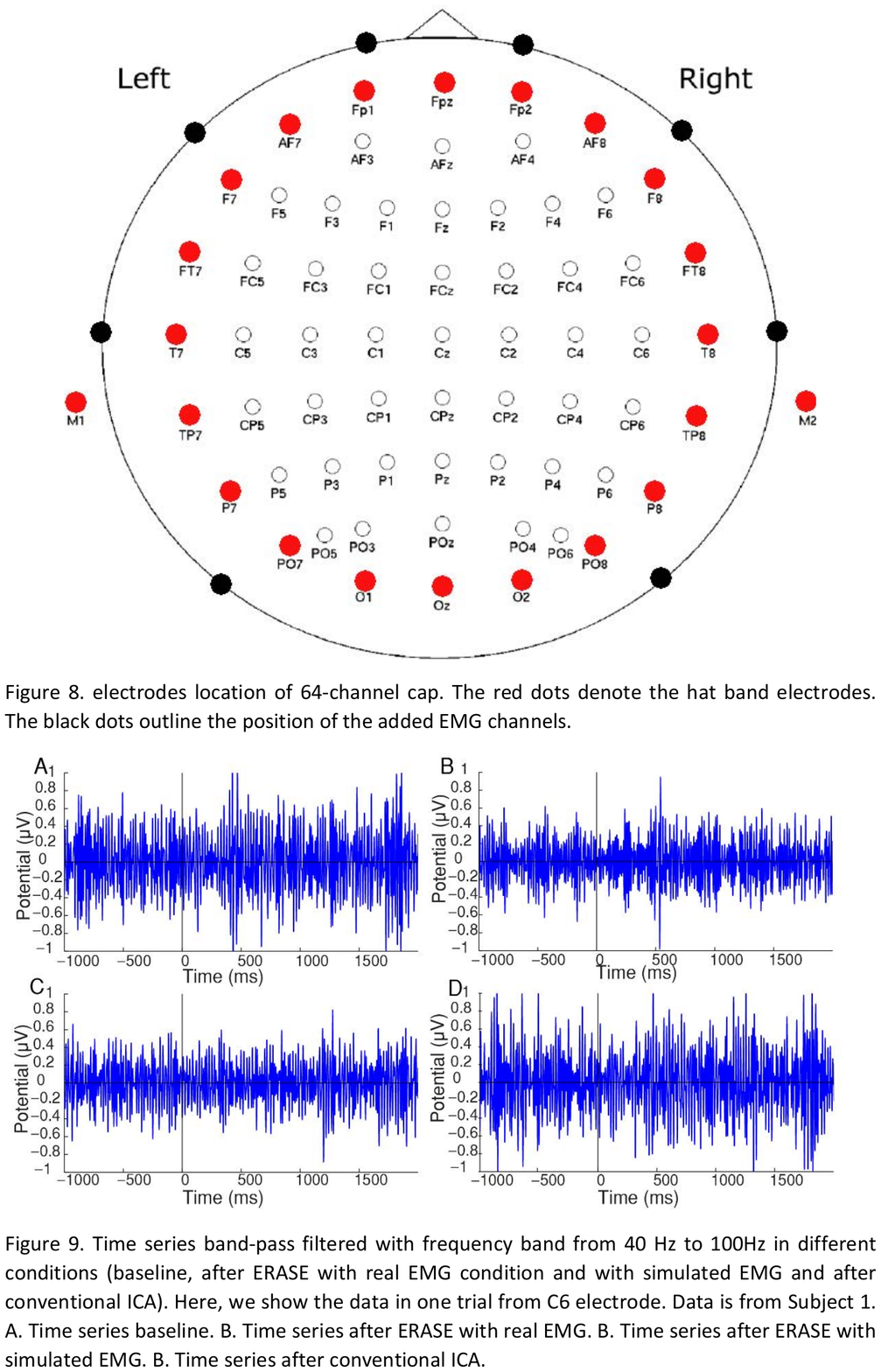} 
\end{figure}

\begin{figure}[!ht]
  \centering
  \includegraphics[width=\textwidth]{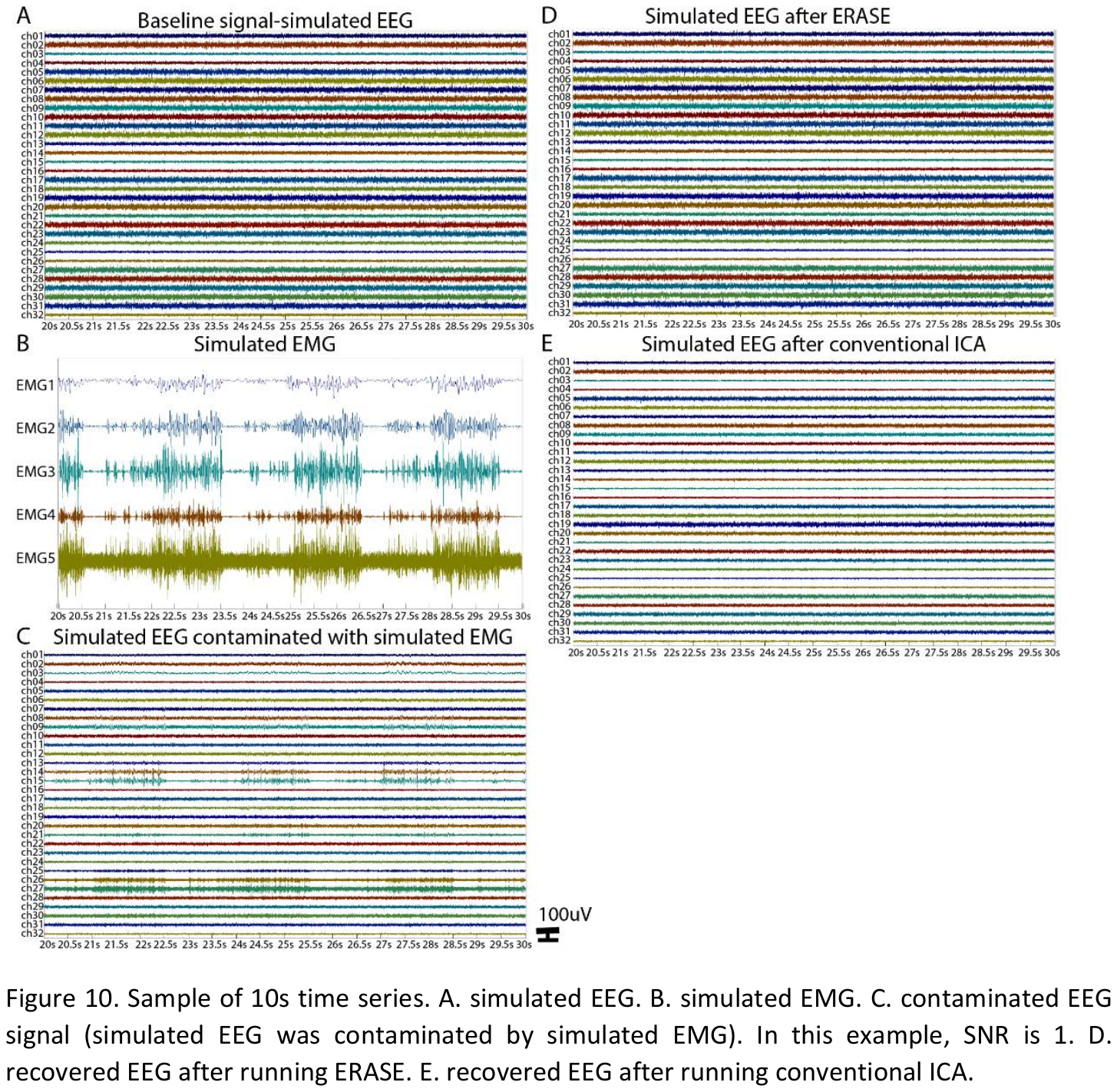} 
\end{figure}

\begin{figure}[!ht]
  \centering
  \includegraphics[width=\textwidth]{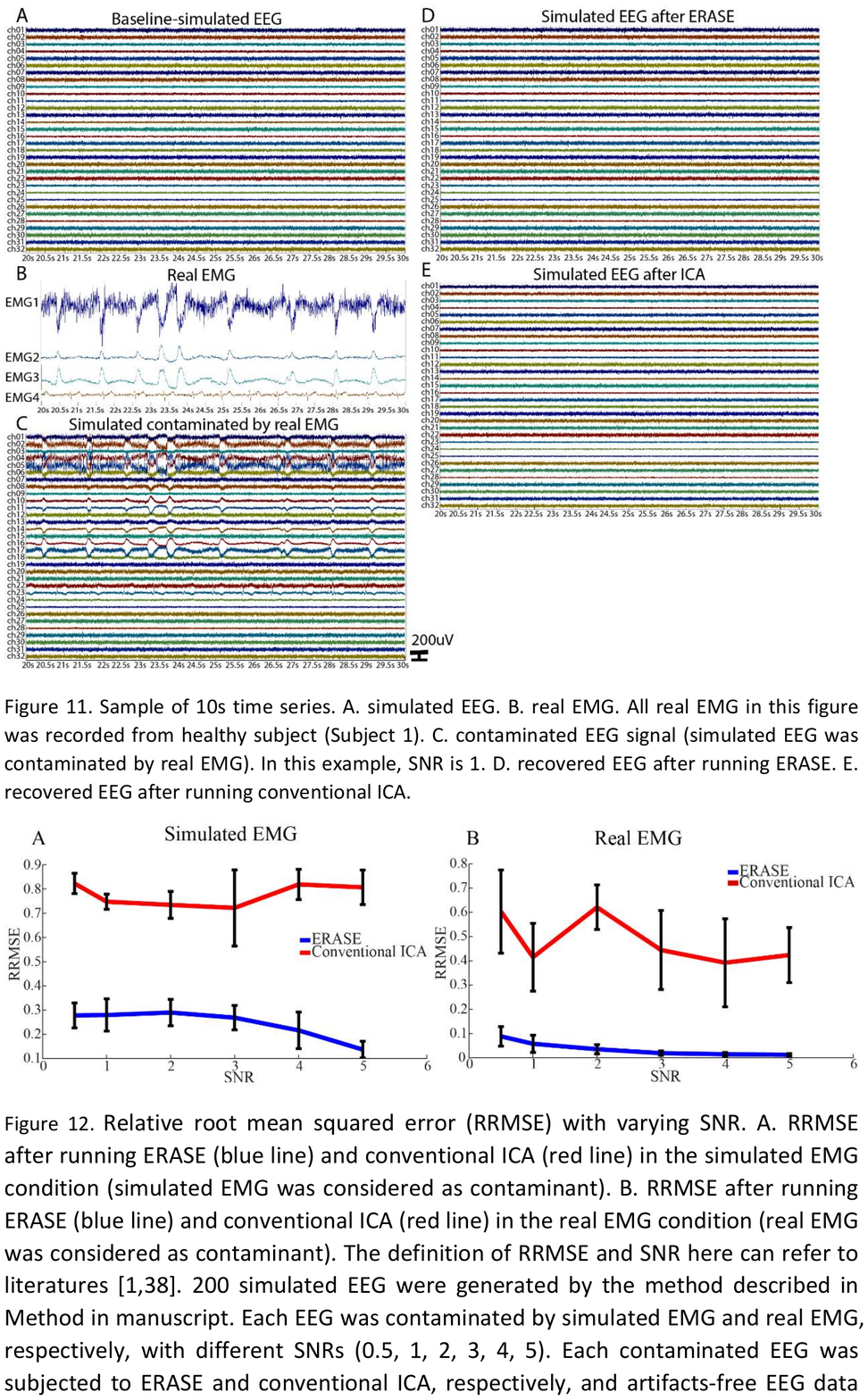} 
\end{figure}

\begin{figure}[!ht]
  \centering
  \includegraphics[width=\textwidth]{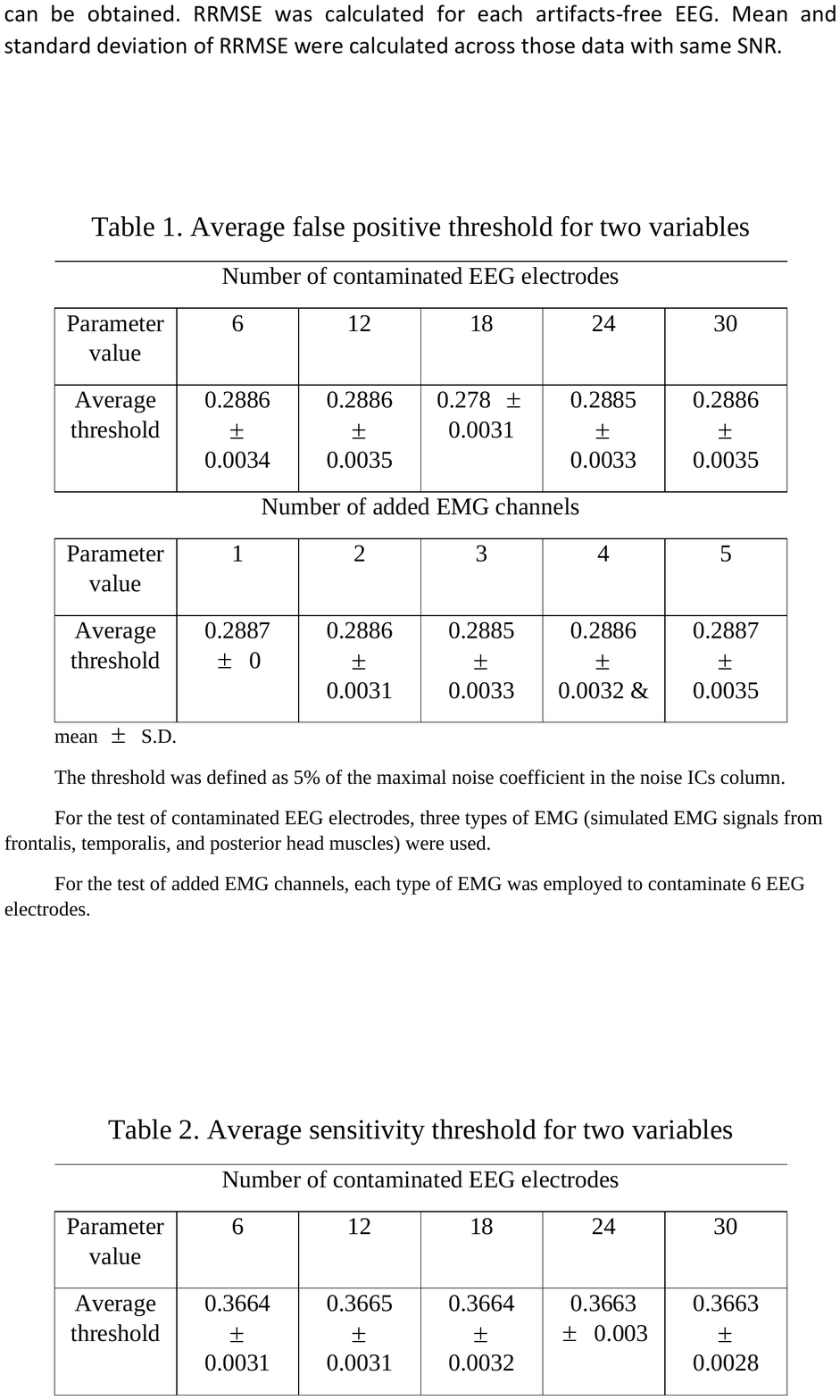} 
\end{figure}

\begin{figure}[!ht]
  \centering
  \includegraphics[width=\textwidth]{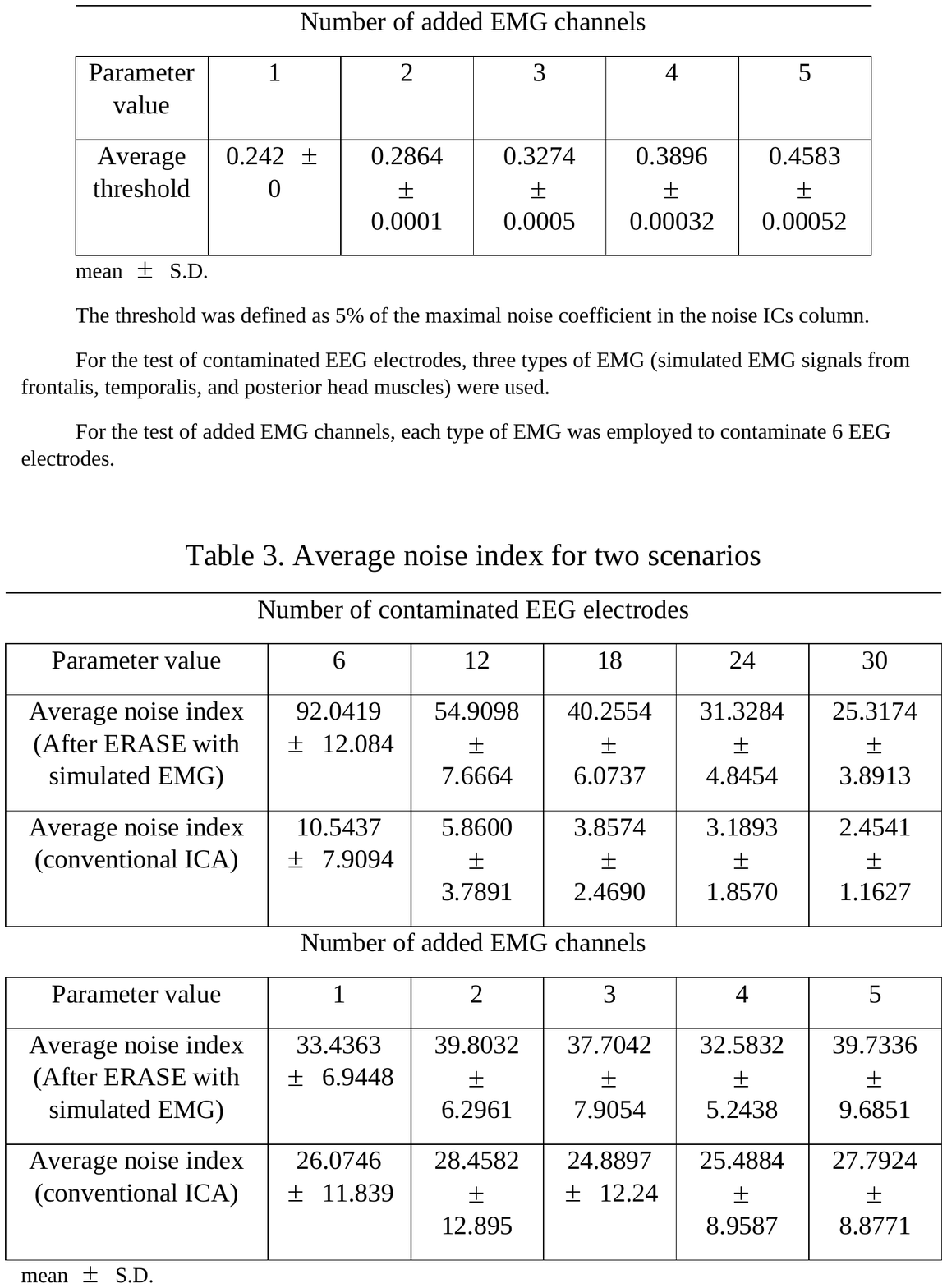} 
\end{figure}

\end{document}